\documentclass[%
 reprint,
 amsmath,amssymb,
 aps,
]{revtex4-2}

\usepackage{graphicx}
\usepackage{tabularx}
\usepackage{subfigure}
\usepackage{upgreek}
\usepackage{url}
\usepackage{amsmath}
\usepackage{siunitx}
\usepackage{dcolumn}
\usepackage{bm}
\usepackage{hyperref}

\makeatletter
\def\url@leostyle{%
  \@ifundefined{selectfont}{\def\UrlFont{\sf}}{\def\UrlFont{\small\ttfamily}}}
\makeatother
\begin{document}

\preprint{APS/123-QED}

\title{A high-resolution, low-latency, bunch-by-bunch feedback system for nano-beam stabilization}

\author{D. R. Bett}
 \altaffiliation[Also at ]{CERN, Geneva, Switzerland}
 
\author{N. Blaskovic Kraljevic}
\altaffiliation[Present address: ]{MAX IV, Lund, Sweden
} 
 
\author{T. Bromwich}
\altaffiliation[Present address: ]{Human Genetics Unit, University of Edinburgh, Edinburgh, United Kingdom
}

\author{P. N. Burrows}

\author{G. B. Christian}
\altaffiliation[Present address: ]{Diamond Light Source, Harwell, United
Kingdom}

 \author{\\C. Perry}

\author{R. Ramjiawan}
 \altaffiliation[Also at ]{CERN, Geneva, Switzerland}
 \email{\\rebecca.ramjiawan@physics.ox.ac.uk}

\affiliation{John Adams Institute for Accelerator Science at University of Oxford,
Denys Wilkinson Building, Keble Road, Oxford, OX1 3RH, United Kingdom}

\date{\today}%

\begin{abstract}
We report the design, operation and performance of a high-resolution, low-latency, bunch-by-bunch feedback system for nano-beam stabilisation. The system employs novel, ultra-low quality-factor cavity beam position monitors (BPMs), a two-stage analogue signal down-mixing system, and a digital signal processing and feedback board incorporating an FPGA. The FPGA firmware allows for the real-time integration of up to fifteen samples of the BPM waveforms within a measured latency of \SI{232}{\ns}. We show that this real-time sample integration improves significantly the beam position resolution and, consequently, the feedback performance. The best demonstrated real-time beam position resolution was \SI{19}{\nm}, which, as far as we are aware, is the best real-time resolution achieved in any operating BPM system. 
The feedback was operated in two complementary modes to stabilise the vertical position of the ultra-small beam produced at the focal point of the ATF2 beamline at KEK. In single-BPM feedback mode, beam stabilization to $50\pm5$~nm was demonstrated. In two-BPM feedback mode, beam stabilization to $41\pm4$~nm was achieved. 
\end{abstract}

\maketitle


\section{INTRODUCTION}

\begin{figure}[htbp]
  \centering
  \includegraphics[width=0.99\linewidth]{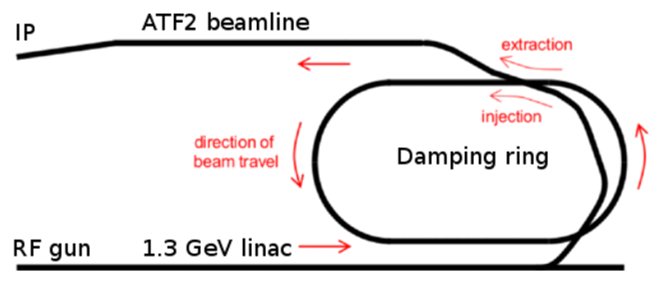}
\caption[Schematic of the ATF2 layout]{Schematic of the ATF layout, with the ATF2 focal point indicated as the IP~\cite{Bett_2021}.}
  \label{fig:ATFDiagram2}
\end{figure}

The Accelerator Test Facility (ATF)~\cite{Aryshev:2742899} is a \SI{1.3}{GeV} electron beamline complex located at the High Energy Research Organisation (KEK) in Tsukuba, Japan. The facility is a test-bed for technologies required for a future linear electron-positron collider. The ATF comprises a linear accelerator, damping ring (DR) and final focus system. Ultra-low emittance beams can be produced with the \SI{138.6}{\m} circumference DR via the process of radiation damping. After extraction from the DR, the beam passes through an extraction line and a final focus system (ATF2~\cite{walker2005atf2, atf22006atf2}) (Fig.~\ref{fig:ATFDiagram2}) which is a scaled prototype of the final focus system of the International Linear Collider (ILC)~\cite{TDR_Vol1} or the Compact Linear Collider (CLIC)~\cite{CLICCDR}. The vertical focal point of the ATF2 beamline is designated the `interaction point' (IP). The two main goals~\cite{bambade2010present, GlenATF} of the ATF2 Collaboration are to produce nano-beams with an IP beam size of 37~nm and position stabilization at the nanometer level.

The ATF is typically operated with an extracted beam pulse repetition rate of 3.12 Hz, a beam charge in the range $0.1-1.0\times10^{10}~e$, and with one bunch per pulse. 
Multi-bunch trains can also be produced by accumulating two or three bunches in the DR and extracting them as a single train with one pulse of the DR extraction kicker.

The Feedback On Nanosecond Timescales (FONT) group~\cite{FONT} has developed several generations of prototype bunch-by-bunch beam-stabilization feedback systems which have been tested at the ATF. A feedback system was deployed in the upstream section of the ATF2 extraction line, using high-resolution bunch-position measurements from stripline beam-position monitors (BPMs)~\cite{PhysRevSTAB.18.032803}, to demonstrate~\cite{PhysRevAccelBeams.21.122802} the resolution, correction-range and latency requirements for the ILC IP beam collision feedback system~\cite{burrows2016fast}. An extended feedback system based on this hardware was recently used to stabilize the beam trajectory before its entrance to the final-focus region, and yielded a significant reduction in the impact of `wakefields' on the beam-size growth~\cite{Bett_2021}.  

\begin{figure}[htbp]
  \centering
  \includegraphics[width=0.99\linewidth]{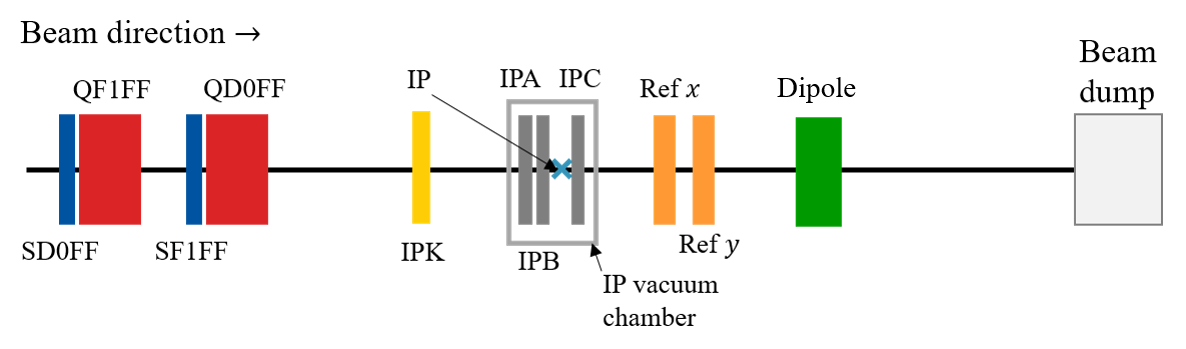}
\caption{Schematic of the ATF2 IP region, showing the final-focus magnets and the elements of the FONT IP feedback system including dipole cavity BPMs IPA, IPB and IPC, reference cavity BPMs Ref $x$ and Ref $y$, and the stripline kicker IPK.}
  \label{fig:ATFDiagram}
\end{figure}

Here we report the design and performance of a high-resolution, high-precision, low-latency, beam-position feedback system located around the ATF2 IP (Fig.~\ref{fig:ATFDiagram}), which is aimed at stabilizing directly the IP vertical beam position to the nanometer level. This system incorporates five cavity BPMs similar to those reported in~\cite{Inoue}, but with a much lower `quality factor'. The down-mixed BPM signals are digitized using a custom FPGA-based feedback controller, the `FONT5A' board~\cite{PhysRevAccelBeams.21.122802, Christian:2312282}, and the feedback calculation is performed on an FPGA mounted on the board. An analogue correction signal is output from the board, amplified using a custom power amplifier with a fast rise-time (35~ns)~\cite{TMD}, and used to drive a stripline kicker, IPK. 

The cavity-BPM system is described in Section II, and its resolution performance is presented in Section III. The bunch-by-bunch feedback system is described in Section IV, and its beam stabilization performance is reported in Section V. A summary of results, and conclusions, is given in Section VI.

\section{IP CAVITY BPM SYSTEM}

The IP BPM system incorporates five C-band cavity BPMs~\cite{jang2013development,kim2012cavity}, IPA, IPB, IPC, Ref $x$ and Ref $y$ (Fig.~\ref{fig:ATFDiagram}). Throughout this paper $x$ and $y$ refer respectively to the horizontal and vertical beam position coordinates in the plane transverse to the beam propagation direction. For beam-size measurements using the IP Beam Size Monitor~\cite{Suehara:2010zz} the IP is placed longitudinally between IPB and IPC. However, for the nano-beam stabilization studies reported here, the IP can instead be placed at any one of IPA, IPB or IPC; this is discussed in Section IV. The cavity BPM design and operation is described below.

\subsection{Cavity BPM design and operation}

As a bunch of charged particles passes through a cavity BPM, its electromagnetic eigenmodes are excited~\cite{Boogert}. The transverse magnetic (TM) modes can be used to determine both the bunch charge and the bunch offset w.r.t.\ the cavity's electrical axis. Separate cavities were designed for sensitivity to the monopole and dipole TM modes, referred to as `reference' and `dipole' cavities respectively; Ref $x$ and Ref $y$ are reference cavities and IPA, IPB and IPC are dipole cavities (Fig.~\ref{fig:ATFDiagram}).

\begin{figure}[htbp]
\centering
\subfigure[]{\includegraphics[width=0.73\linewidth]{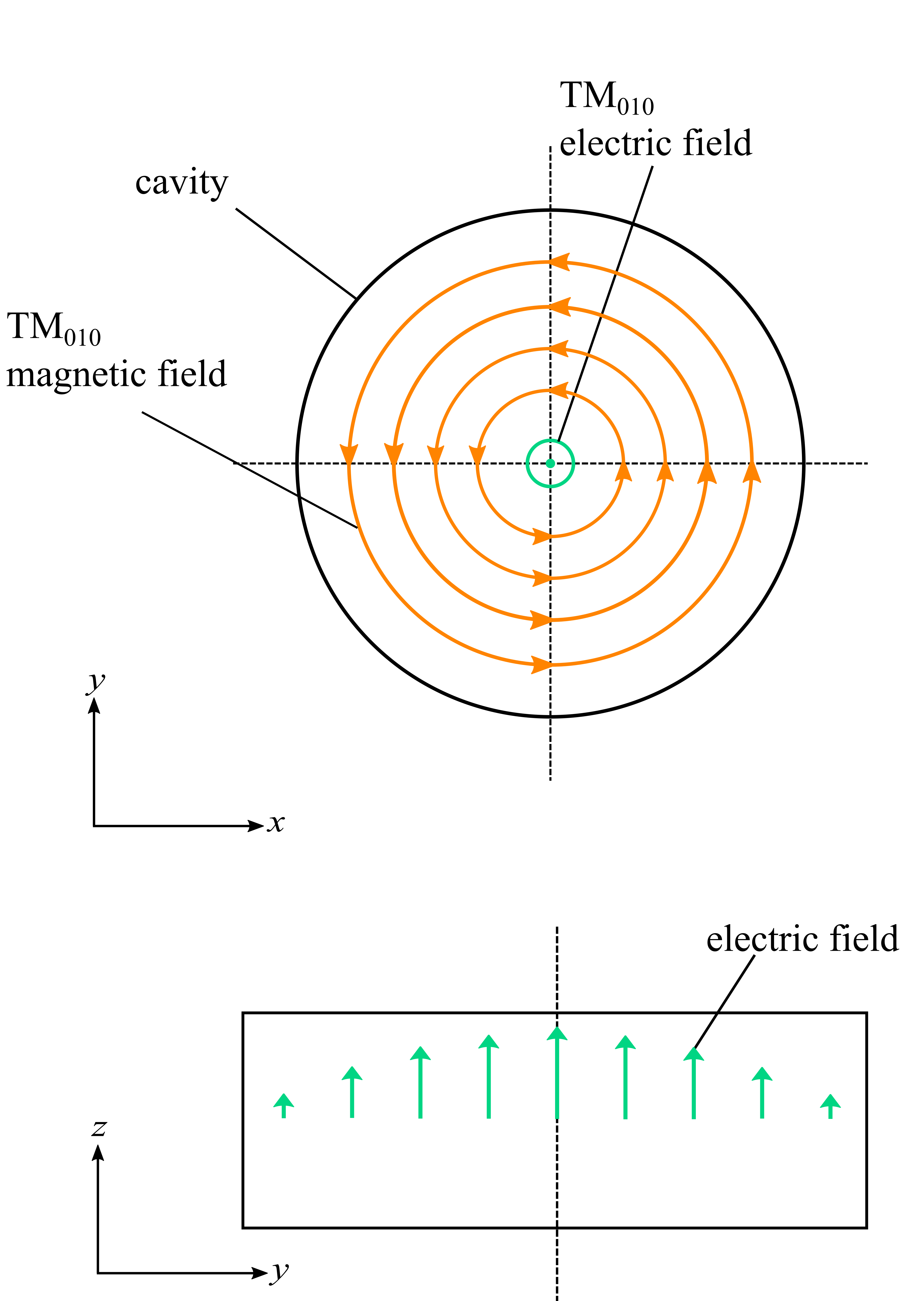}}
\hfill
\subfigure[]{\includegraphics[width=0.73\linewidth]{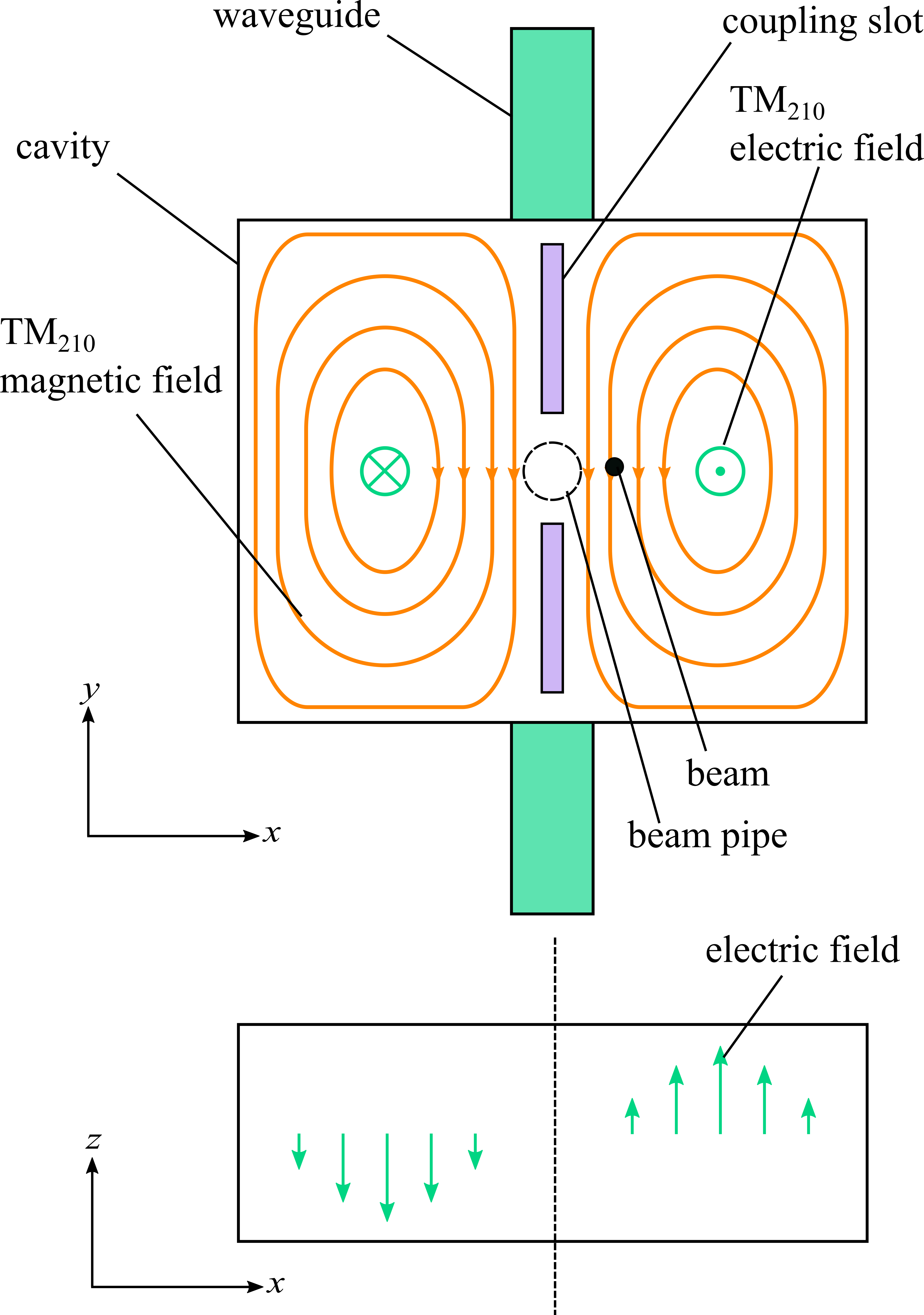}}
\caption{Schematic of the electric and magnetic field lines of (a) the TM010 mode for
a circular cylindrical cavity BPM and (b) the TM210 mode (or $x$-dipole mode) for a rectangular cylindrical cavity BPM. The
waveguides which couple to the TM210 mode are shown. There are also corresponding waveguides which couple to the $y$-dipole TM120 mode.}
\label{fig:Extraction}
\end{figure}

\begin{figure}[htbp]
  \centering  
\subfigure[]{\includegraphics[width=0.4\textwidth]{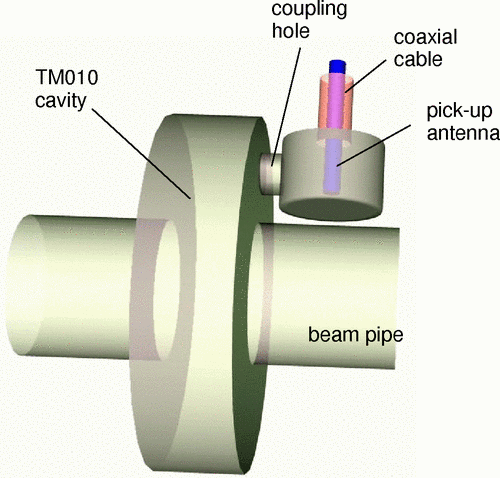}{}}\hfill
\subfigure[]{\includegraphics[width=0.4\textwidth]{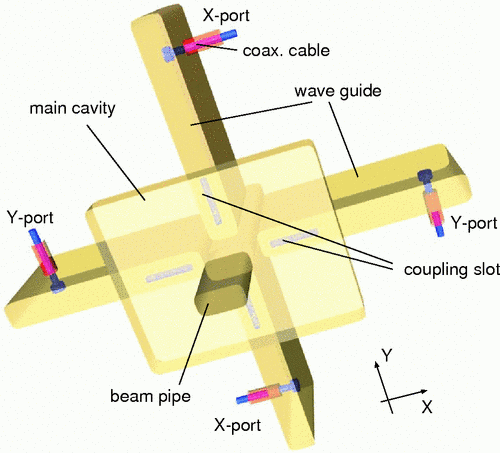} }\hfill
  \caption{Schematic of the (a) cylindrical reference cavity BPMs and (b) rectangular dipole
cavity BPMs~\cite{Inoue}.}
\label{fig:CavityDiagram}
\end{figure}

In the circular cylindrical $x$ and $y$ reference cavities the dominant excited mode is the monopole mode, illustrated in Fig.~\ref{fig:Extraction}(a), which is sensitive to the bunch charge and, for small offsets, insensitive to its position offset w.r.t.\ the cavity electrical center. A schematic of the reference cavity is shown in Fig.~\ref{fig:CavityDiagram}(a), with the coupling slot and antenna indicated~\cite{NevenThesis}. The $x$ and $y$ reference cavities have diameters of 42.95 and 38.65~mm respectively, designed to yield monopole-mode frequencies equal to the respective dipole-mode frequencies of the dipole cavities (see below). Using dedicated tuning pins, the monopole-mode frequency of each reference cavity was fine-tuned to match the respective dipole-mode frequency of the dipole cavities (see Table~\ref{CavityFrequencies}).

The dipole cavity principle is illustrated in Fig.~\ref{fig:Extraction}(b). The cavity design is of rectangular cylinder form and uses spatial filtering to suppress the dominant monopole mode so that the higher-frequency dipole mode can be extracted~\cite{SelectiveCoupling}; this mode is sensitive to the bunch position offset as well as its charge. A schematic of the cavities used is shown in Fig.~\ref{fig:CavityDiagram}(b), with the coupling slots and waveguides indicated. The cavities were designed with different vertical and horizontal dimensions so as to decouple the horizontal and vertical dipole modes; the positioning of the coupling slots and waveguides allows these modes to be extracted separately from the same BPM. There are pairs of output $x$-ports and $y$-ports in each cavity; the respective output signals are combined (see Fig.~\ref{fig:4ProcessingElectronics}) so as to double the signal from the anti-symmetric dipole mode and cancel the unwanted symmetric monopole mode. A \SI{700}{MHz} bandwidth band-pass filter (BPF) removes the residual monopole signal.

The cavities are fabricated from aluminium and were designed~\cite{jang2013development} to have ultra-low quality-factor values so as to be suitable for resolving in time individual particle bunches in trains with bunch separations of order 100~ns. The design and measured values of the cavity resonant frequencies are given in Table~\ref{CavityFrequencies}. Also given are the measured signal exponential decay times, which are around 25~ns for the dipole cavities and 14~ns for the reference cavities. 

\begin{table}[htbp]
\begin{center}
\centering
\newcolumntype{Y}{>{\centering\arraybackslash\hsize=0.3\hsize}X}
\newcolumntype{Z}{>{\centering\arraybackslash\hsize=0.8\hsize}X}
\begin{tabularx}{\linewidth}{Z Y Y Y}\toprule
BPM cavity &  Design& Measured & Decay\\ 
& frequency & frequency & time \\
 & (GHz) & (GHz) & (ns)\\\hline\\[-0.9em]
Dipole IPA ($x$-port) & 5.712 & 5.705 &  25\\
Dipole IPB ($x$-port) & 5.712 & 5.706 &  25\\
Dipole IPC ($x$-port) & 5.712 & 5.704 &  23\\\hline\\[-0.9em]
Dipole IPA ($y$-port) & 6.426 & 6.428 &  26\\
Dipole IPB ($y$-port) & 6.426 & 6.427 &  22\\
Dipole IPC ($y$-port) & 6.426 & 6.428 &  21\\\hline\\[-0.9em]
Reference ($x$-cavity) & 5.711 & 5.705 & 14\\
Reference ($y$-cavity) & 6.415  & 6.428 & 14\\
\toprule
\end{tabularx}
\end{center}
\caption[Resonant frequencies of the dipole and reference cavity BPMs]{Design and measured values of the resonant frequencies of the dipole~\cite{jang2013development} and reference~\cite{TalithaThesis} cavity BPMs, and measured signal decay times.}
\label{CavityFrequencies}
\end{table}

Since the ATF2 is designed to focus the beam to c. 37~nm in the vertical ($y$) plane, and our aim is nano-beam stabilization in this plane, for the remainder of this paper we consider only vertical beam position measurements and hence discuss only those signals from the reference $y$-cavity BPM (Ref $y$) and the dipole BPM (IPA, IPB, IPC) $y$-ports.

The position resolution of a dipole cavity BPM is primarily limited by the signal-to-noise ratio, where the signal level is determined both by how much energy is transferred from the beam to the dipole modes and also how well this mode is coupled out of the BPM through the waveguides. Sources of noise in the system include thermal and electronic noise, as well as signal contamination from the monopole mode~\cite{Boogert}. The resolution performance is discussed in Section~\ref{sec:resolution}.

A variable attenuator on the combined output of each dipole BPM (Fig.~\ref{fig:4ProcessingElectronics}) can be used to increase the dynamic range of the position measurement but at the expense of the resolution. For typical operating bunch charges of 1~nC, 10~dB attenuation was added to the dipole signal, yielding a dynamic range for vertical position measurements of~\SI{\pm 3}{\um}.

\begin{figure}[hbtp]
  \centering
  \includegraphics[width=0.98\linewidth]{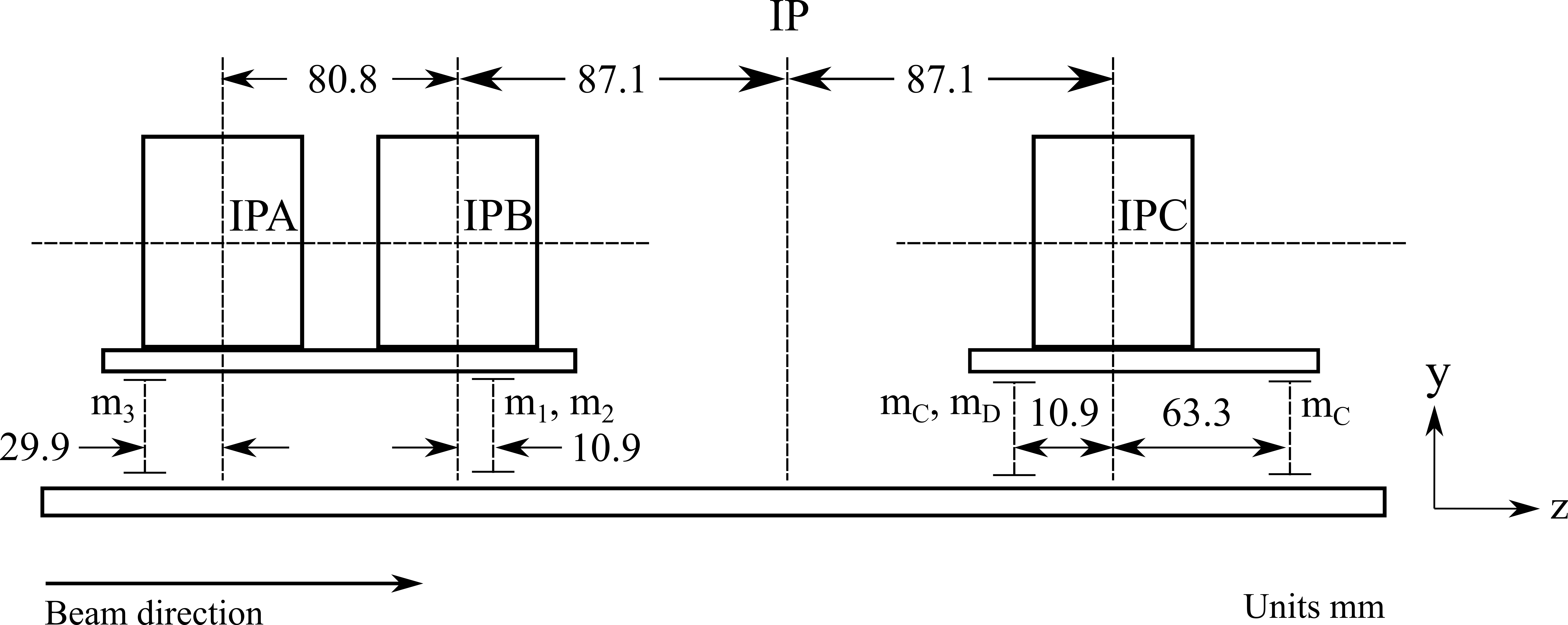}
\caption{Schematic of the IP BPM configuration, showing the `IPAB' mover block, with submovers $\mathrm{m_1}$, $\mathrm{m_2}$ and $\mathrm{m_3}$, on which IPA and IPB are mounted, and the IPC mover block with submovers $\mathrm{m_C}$, $\mathrm{m_D}$ and $\mathrm{m_E}$. The nominal IP location, as used for beam-size measurements, is indicated.}
  \label{fig:Submovers}
\end{figure}

The dipole cavity BPMs are mounted on two piezo-mover systems (Fig.~\ref{fig:Submovers}) within the vacuum chamber which allow horizontal, vertical and angular BPM alignments w.r.t.\ the beam trajectory~\cite{Blanco}. IPA and IPB are mounted on a single `IPAB' mover block and, therefore, cannot be moved independently. The IPAB movers were manufactured by Cedrat Technologies and have a working range of \SI{248}{\um}, while the IPC mover was manufactured by PI and has a working range of \SI{300}{\um}. The movers incorporate feedback systems designed to ensure a position stability of better than 2~nm.

\subsection{BPM analogue signal processing}

\begin{figure*}[htbp]
\centering
\includegraphics[width=0.9\textwidth]{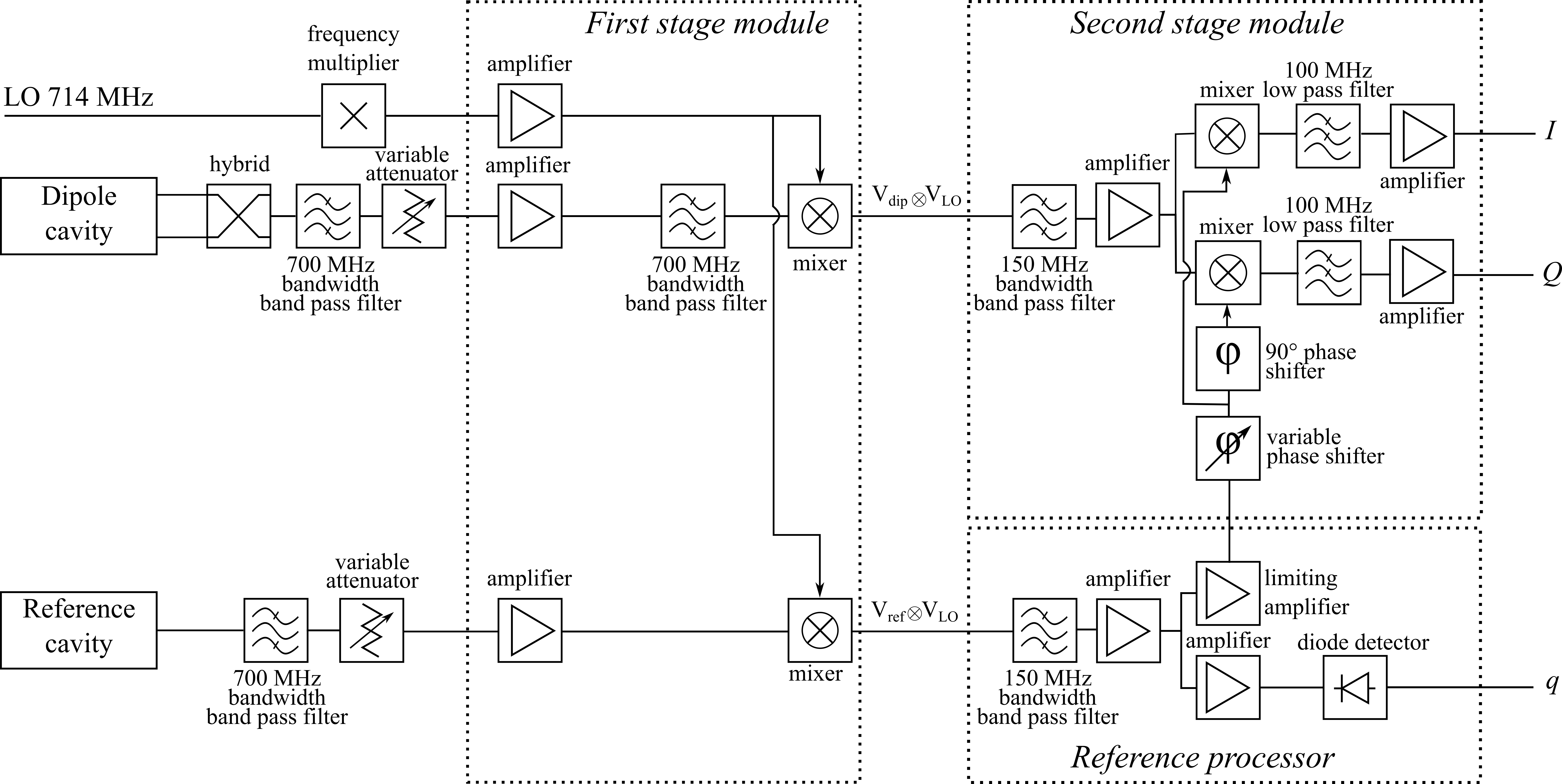}
\caption[Simplified block diagram of two-stage processing electronics]{Simplified block diagram of the two-stage down-mixing process of the dipole and reference cavity signals from GHz-level to baseband. Diagram adapted from~\cite{kim2012cavity}.}
\label{fig:4ProcessingElectronics}
\end{figure*}

The cavity BPM signals undergo two stages of frequency down-mixing~\cite{Inoue, NevenThesis} (see Fig.~\ref{fig:4ProcessingElectronics}) so as to produce baseband signals that can be digitized with the FONT5A board. In the first stage, both the reference and dipole cavity signals are down-mixed to an intermediate frequency (IF) centered at 714~MHz using a common Local Oscillator (LO) signal so as to retain the phase relation between the signals. The 5.712~GHz LO signal is generated using frequency multiplication of the Master Oscillator signal~\cite{TimingSystem} and hence is phase-locked to the beam. The LO signal can be written
\begin{equation}\label{VLOInput}
V_{LO} \sim L\sin (2\pi f_{\textup{LO}}t+\Delta\phi_{\textup{LO}}),
\end{equation}
where $f_{\textup{LO}}$ = 5.712~GHz, $\Delta\phi_{\textup{LO}}$ is the phase difference between the LO signal and the dipole signal, and $L$ is a constant.

The output signals from the $y$-port of a dipole cavity ($V_{\textup{dip}}$) and from the $y$ reference cavity ($V_{\textup{ref}}$) can be represented by:
\begin{subequations}
\begin{equation} \label{VdipInput} 
V_{\textup{dip}}  \sim q(D_{y}y\sin (2\pi f_{\textup{dip}}t) +(D_{y'}y'-D_{\alpha}\alpha)\cos(2\pi f_{\textup{dip}}t))
\end{equation}
and
\begin{equation}\label{VrefInput}
V_{\textup{ref}}  \sim q(\sin (2\pi f_{\textup{ref}}t+\Delta\phi)),
\end{equation}
\end{subequations}
where $q$ is the bunch charge, $y$ is the vertical beam position offset w.r.t.\ the cavity electrical axis, $y'$ and $\alpha$ are the bunch pitch-angle and angle-of-attack, respectively, $f_{\textup{dip}}$ or $f_{\textup{ref}}$ is the respective signal frequency, and $\Delta\phi$ is the difference in phase between the monopole and dipole signals; $D_y$, $D_{y'}$ and $D_{\alpha}$ are constants. It can be seen that the signals excited by a $y'$ or $\alpha$ offset are \ang{90} out of phase with those excited by a $y$ offset. For small bunch offsets the reference-cavity signal is independent of the beam position.  

After the first stage of down-mixing, signals $V_{\textup{dip}}\otimes V_{\textup{LO}}$ and  $V_{\textup{ref}}\otimes V_{\textup{LO}}$ are produced (Fig.~\ref{fig:4ProcessingElectronics}) at both the IF (714~MHz) and 
 the higher frequencies $f_{\textup{dip}}+ f_{\textup{LO}}$ or $f_{\textup{ref}}+ f_{\textup{LO}}$, respectively. 
 In the second-stage processing, the latter are removed with a \SI{150}{MHz} bandwidth band-pass filter centerd at c. 700 MHz. The $V_{\textup{ref}}\otimes V_{\textup{LO}}$ IF signal is then split, with one of the outputs passing through a diode detector to produce a pulse whose magnitude is proportional to the bunch charge. This signal, subsequently denoted $q$, is used for bunch-charge normalization to obtain the bunch position (see below). The other $V_{\textup{ref}}\otimes V_{\textup{LO}}$ output passes through a limiting amplifier to remove its charge dependence. This signal is then used as the LO signal for the second stage of down-mixing of the dipole signals, from the IF to baseband.

In the second stage, the reference and dipole signals are mixed in-phase and in-quadrature to produce $I$ and $Q$ signals, respectively. These signals are orthogonal components that together include the full amplitude and phase information of the BPM waveform~\cite{IPACCavityNeven}. If the BPM is well-aligned in $y'$ and $\alpha$, the contributions to $V_{\textup{dip}}$ from these terms are much smaller than those from $y$, such that
\begin{equation}\label{VI}
\begin{split}
I&=(V_{\textup{dip}}\otimes V_{\textup{LO}})\otimes (V_{\textup{ref}}\otimes V_{\textup{LO}})\\
&\propto qy\cos(\theta_{IQ}),
\end{split}
\end{equation}
and
\begin{equation}\label{VQReduced}
Q \propto qy\sin(\theta_{IQ}),
\end{equation}
where the phase angle, $\theta_{IQ}$, corresponds to
\begin{equation}\label{DefineTheta1}
\theta_{IQ}=2\pi (f_{\textup{dip}}-f_{\textup{ref}})t-\Delta\phi.
\end{equation}
Since the reference cavity is tuned such that $f_{\textup{ref}}\simeq f_{\textup{dip}}$ (see Table~\ref{CavityFrequencies}), the $I$ and $Q$ signals are at baseband. Before digitization these signals are amplified so as to reduce the effect of quantization noise.

\subsection{Signal digitization and digital processing}
\begin{figure}[htbp]
\centering
\includegraphics[width=0.48\textwidth]{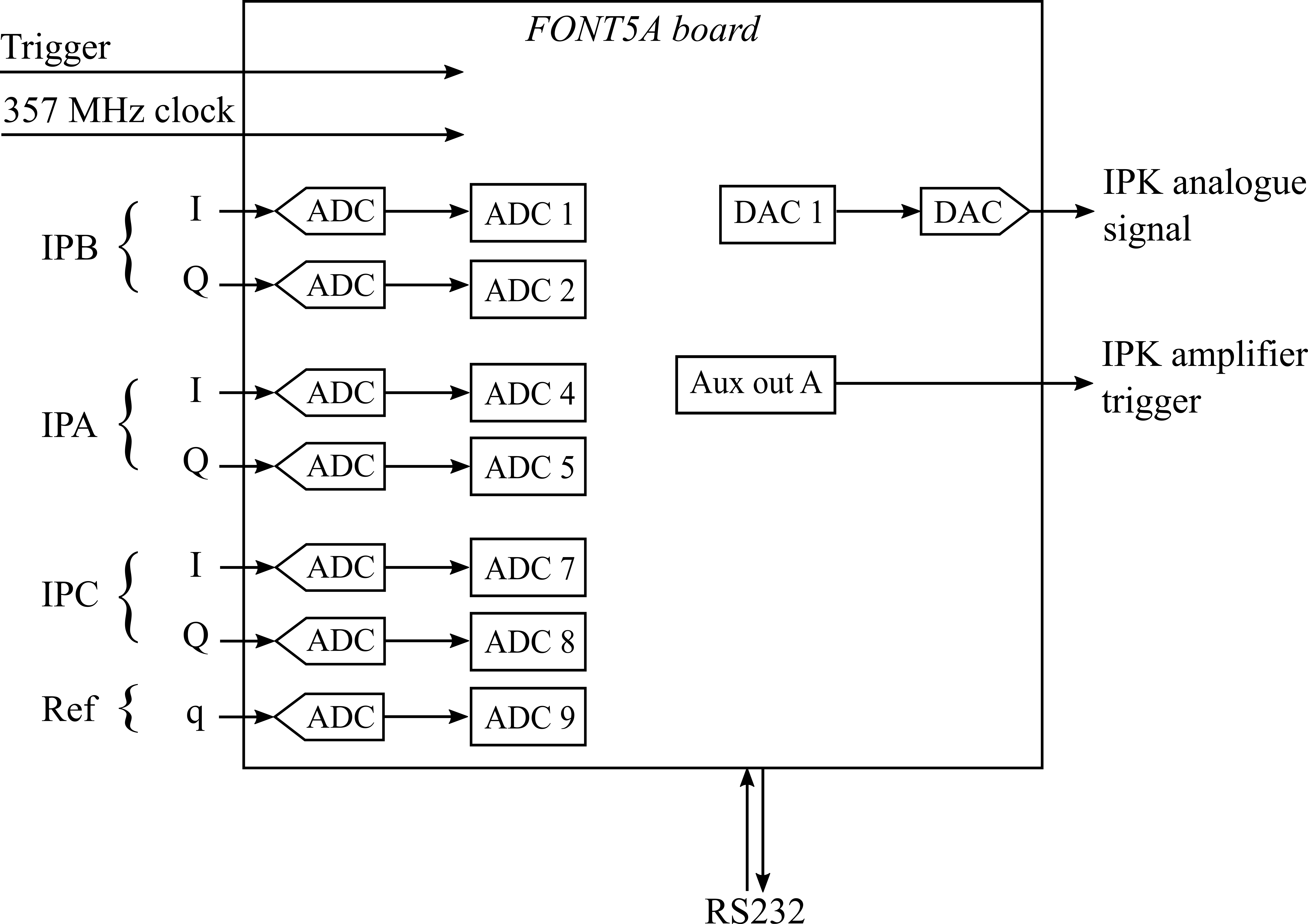}
\caption{Block diagram of the FONT5A digital board showing the primary input and output signals used.}
\label{fig:4FONTBoard}
\end{figure}

\begin{figure}[htbp]
\centering
\includegraphics[width=0.4\textwidth,trim=1.0cm 0cm 1.0cm 0.8cm,clip]{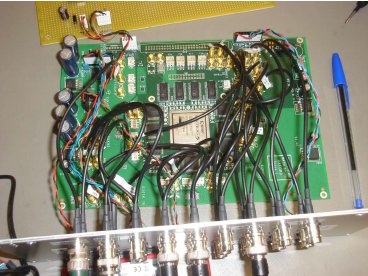}
\caption[Photograph of the FONT5A digital board]{Photograph of the FONT5A digital board with the case removed~\cite{ConstanceThesis,christian2011latest}.}
\label{fig:FONTBoardPhoto}
\end{figure}

The signal digitization is performed on a FONT5A board~\cite{PhysRevAccelBeams.21.122802}, a custom feedback controller with a Xilinx Virtex-5 XC5VLX50T FPGA at its core~\cite{Virtex5}. The primary inputs and outputs of this board are shown in Fig.~\ref{fig:4FONTBoard}. The FPGA firmware is written in the Verilog hardware description language, and the configuration bitstream is stored on a non-volatile Xilinx XCF32P Programmable Read-Only Memory (PROM), from where it is loaded on power-up or system reset. The inputs and outputs of the PCB are via Micro-Coaxial connectors (MCX) which patch to BNC connectors on the case which houses the board~\cite{ConstanceThesis}. The FONT5A board is shown in Fig.~\ref{fig:FONTBoardPhoto} with the case removed. 

The board contains nine Texas Instruments 14-bit ADS5474~\cite{TexasInstruments} analogue-to-digital converters (ADCs) grouped into separately-clocked banks of three. Seven ADCs are used to digitize the $I$ and $Q$ waveforms from IPA, IPB and IPC, and the $q$ waveform from the Ref~$y$ cavity; the least significant bit is removed as it corresponds to the noise level of the signals~\cite{ConstanceThesis}. The ADC channels contain an inherent offset on their baseline signal which can be zeroed by coupling each with the output of a 16-bit DAC, referred to as a trim DAC~\cite{BettThesis}. The values used for the trim DAC  can be set using the associated FONT LabVIEW DAQ which is used to transmit values to the board through an RS-232 Universal Asynchronous Receiver/Transmitter (UART) via an Ethernet serial device server.

\begin{figure}[htbp]
\centering
\includegraphics[width=0.48\textwidth]{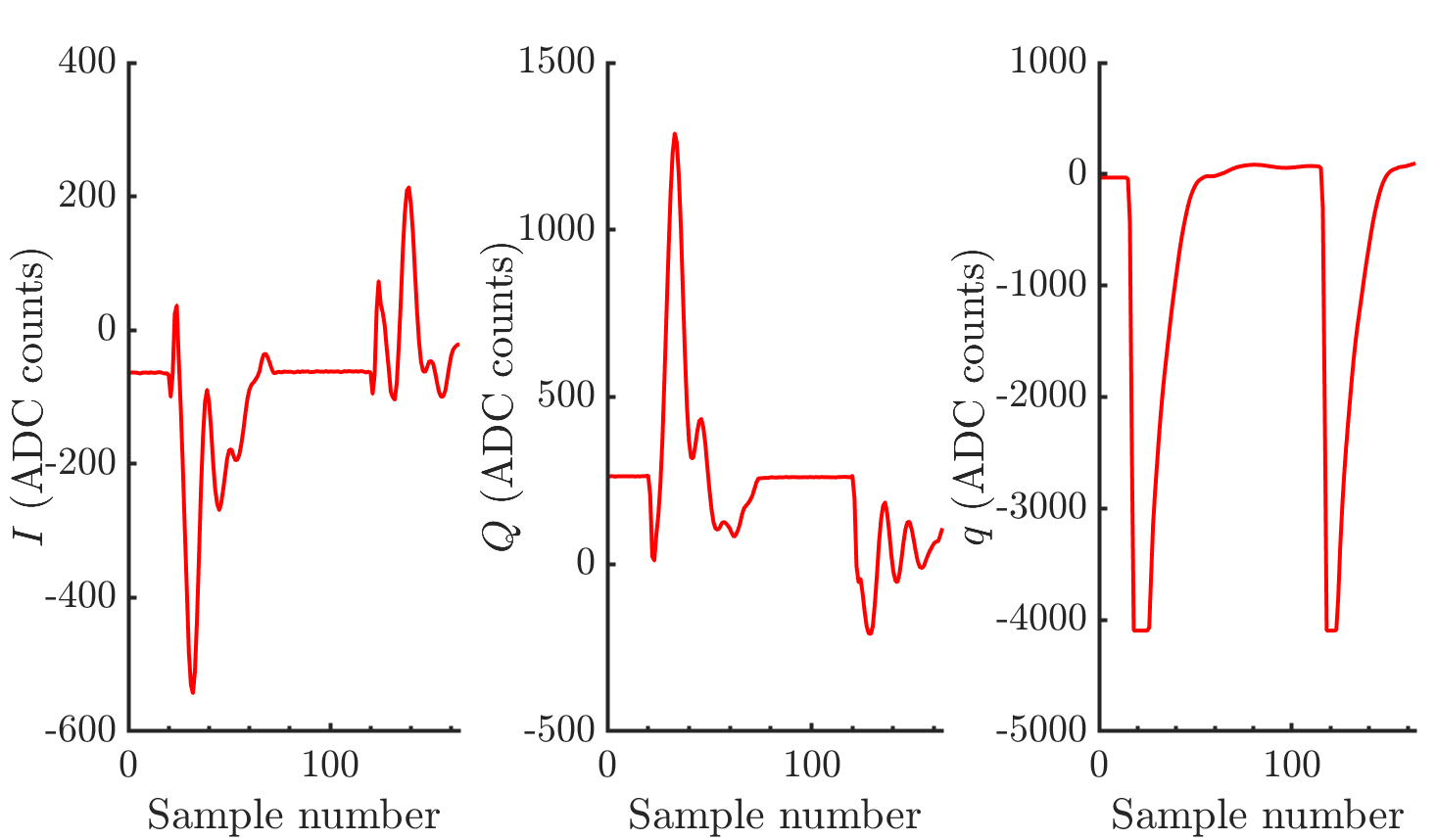}
\caption{Representative digitized $I$, $Q$ and $q$ waveforms from IPC, for two-bunch-train operation with a bunch spacing of 280~ns. The waveforms were sampled at intervals of 2.8~ns.}
\label{fig:IQWaveforms}
\end{figure}

A clock at \SI{357}{\MHz} is used for the time-critical FPGA logic. It is derived from the LO, meaning it is phase-locked to the beam, and used to clock the ADCs, so that the $I$, $Q$ and $q$ signals are digitized at 357~MHz. The start of the sampling window is set with respect to the trigger, which is internally delayed on the board and can be adjusted. The sampling window can be varied within the firmware but for one- or two-bunch operation typically consists of 164 samples each separated by 2.8~ns, meaning that a complete DR beam-circulation period (462~ns) can be digitized within a single window. Representative digitized waveforms for 2-bunch-train operation (Section~\ref{sec:accel}) are shown in Fig.~\ref{fig:IQWaveforms}. The difference in the $I$ and $Q$ signals between the two bunches derives from the transverse position offset between them. 

The firmware includes the functionality to provide a constant offset to the $I$, $Q$ and $q$ signals before they are used to calculate the bunch position. This is used to remove the position-independent baseline signals that are generated on each $I$ and $Q$ waveform at the second stage of the signal processing (Fig.~\ref{fig:IQWaveforms}). 

\subsection{Position measurement}
\label{sec:position}

A linear combination of $I$ and $Q$ can be chosen to produce a signal, $I'$, with an amplitude proportional to the bunch position $y$~\cite{Inoue}:
\begin{equation}\label{DefineVIPrime}
I'=I\cos(\theta_{IQ})+Q\sin(\theta_{IQ}).
\end{equation}
By substituting Eqs.~\ref{VI} and~\ref{VQReduced} into Eq.~\ref{DefineVIPrime}, 
\begin{equation}
\label{AfterExpPrimed2}
y=\frac{1}{k}\frac{I'}{q},
\end{equation}
where $k$ [$\SI{}{\um}^{-1}$] is a constant, found by calibration of the BPM.  
A signal orthogonal to $I'$ can also be generated, $Q'$, that is proportional to the beam pitch: 
\begin{equation}\label{DefineVQPrime}
Q'=-I\sin(\theta_{IQ})+Q\cos(\theta_{IQ}).
\end{equation}

Each dipole BPM is calibrated w.r.t.\ position by vertically scanning the beam across a known range by changing the position of quadrupole QD0FF (Fig.~\ref{fig:ATFDiagram}) and measuring the corresponding BPM response. Calibrations w.r.t.\ the beam angle $y'$ are performed by tilting the BPMs through a known range using the submovers shown in Fig.~\ref{fig:Submovers}. 

For each measured bunch in the beam, the calibration calculation can be performed using either single or multiple samples of the $I$ and $Q$ waveforms. For convenience a single sample of the $q$ signal from Ref $y$ is used for charge normalization of the $I'$ and $Q'$ signals from all three dipole BPMs. The requirements for low-latency feedback preclude the direct implementation of division for the charge normalization within the firmware and, instead, a method of lookup tables (LUTs) is employed using block RAM resources in the FPGA. The charge, $q$, is used as an address to the LUTs, for which the elements are preloaded with $\frac{1}{q}$ scaled by the appropriate feedback coefficient $C_i$,
\begin{equation}\label{LUTCalc}
q\xrightarrow{\text{$LUT_i$}} \frac{C_i}{q},
\end{equation}
where $C_i$ incorporates the terms involving $\theta_{IQ}$, $k$ and the feedback gain $G$ (see Section~\ref{feedback_calculation}); there are four instances of the LUT logic ($1\leq i\leq 4$), each loaded with the respective value of $G$, allowing for up to any two of the BPMs to be used as input to the feedback system~\cite{RamjiawanThesis}.

\begin{figure*}[htbp]
  \centering  
\subfigure[]{\includegraphics[width=0.30\textwidth]{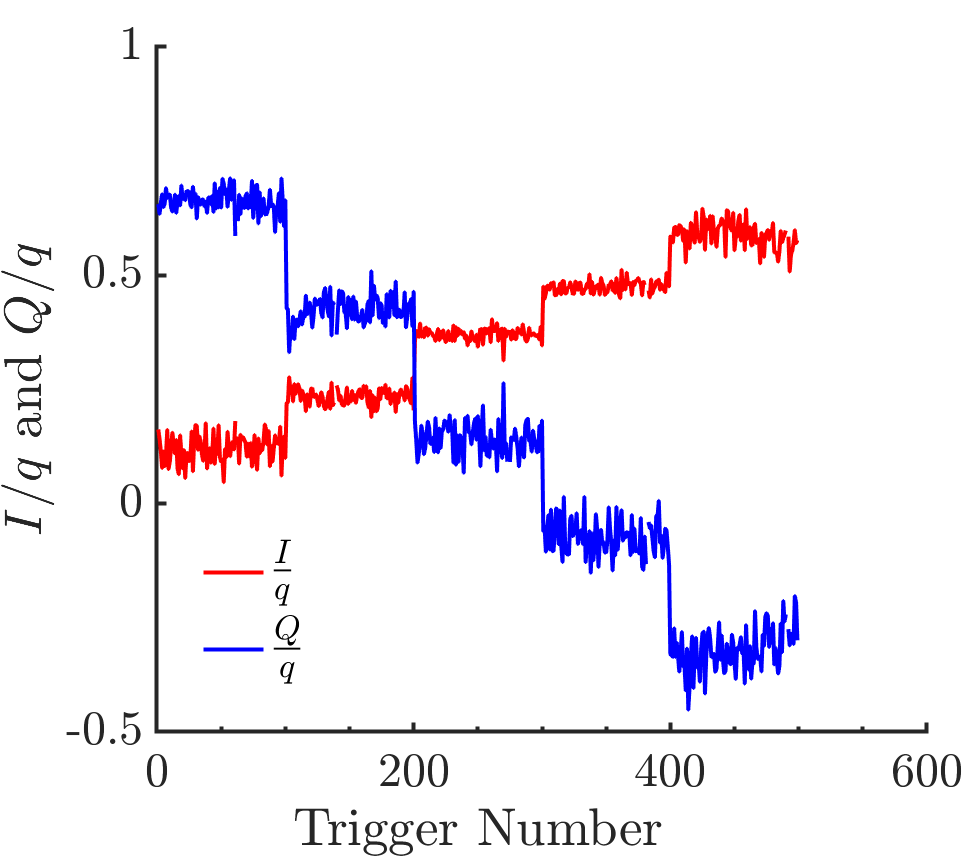}\label{fig:f1IPACalAQD}}
\subfigure[]{\includegraphics[width=0.3\textwidth]{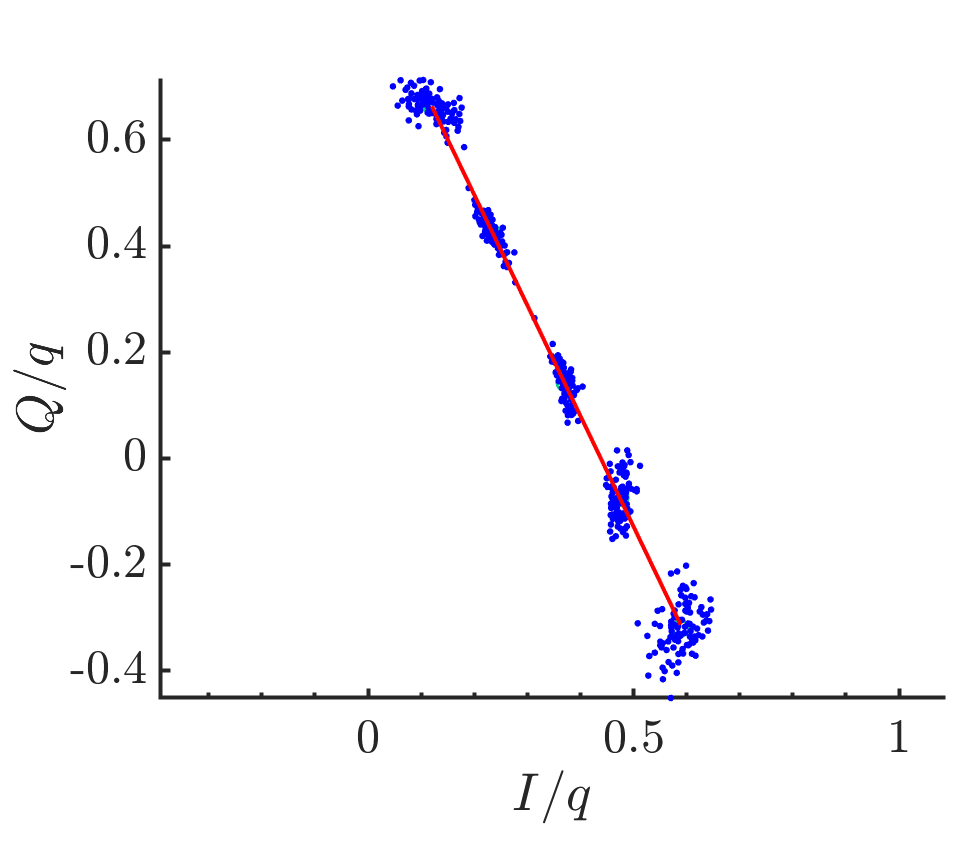}{ }\label{fig:f2IPACalIQAQD}}
 \subfigure[]{\includegraphics[width=0.30\textwidth]{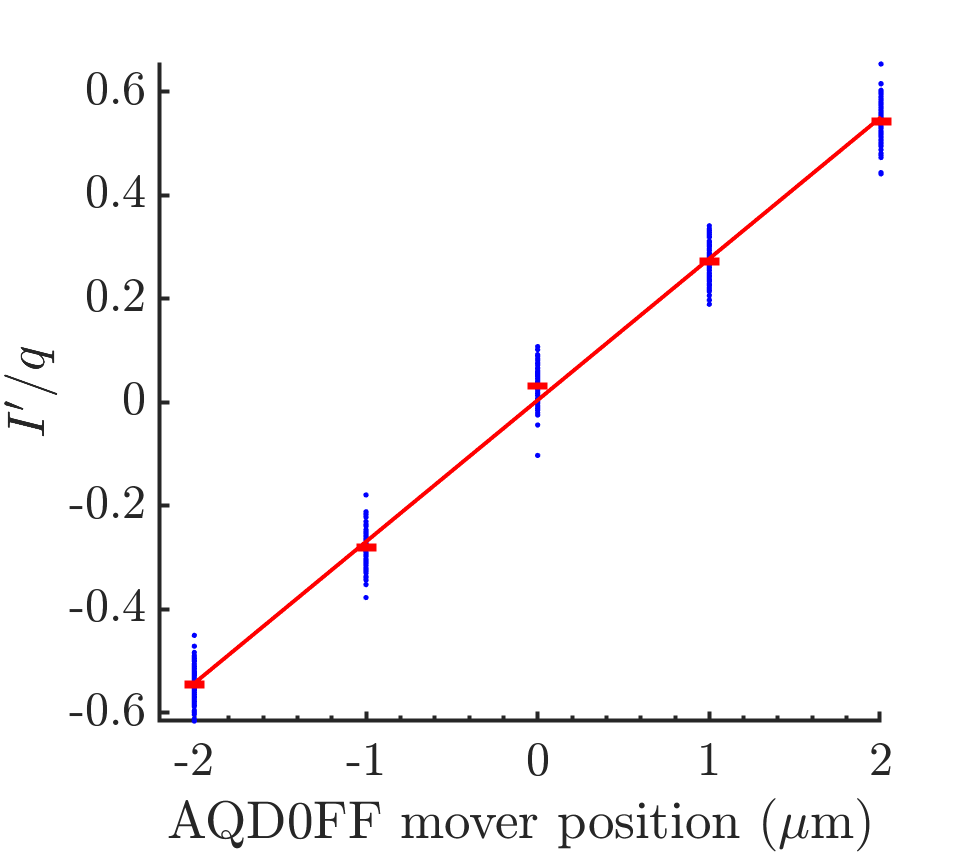}\label{fig:f3IPACalkAQD}}

  \caption[IPA position calibration]{Example vertical position calibration of IPA, using an 11-sample integration range: (a) $\frac{I}{q}$ and $\frac{Q}{q}$ versus trigger number; (b) ${\frac{Q}{q}~\mathrm{versus}~\frac{I}{q}}$ (points); the line shows a least-squares fit to determine ${\theta_{IQ} = -1.093\pm0.006~\mathrm{radians}}$; (c) the data points show $\frac{I'}{q}$ versus QD0FF mover position, the red error bars show the standard error on the mean values at each QD0FF setting and the red line shows a least-squares fit, which yields ${k = 0.184 \pm 0.002~\SI{}{\um^{-1}}}$.}
\label{fig:f4IPA1CalQPrime}
\end{figure*}

The position resolution can be significantly improved (see Section~\ref{sec:resolution}) by integrating over multiple samples of the $I$ and $Q$ signals as this both increases the signal level and averages over thermal and electronic noise. The integration range is chosen around the peak of the $I$ and $Q$ signals, as samples significantly in advance of the peak may contain transient effects from unwanted modes and samples late in the waveform have a poorer signal-to-noise ratio. This integration is performed in real time on the FONT5A board: on every rising fast-clock edge within the selected integration window, the most recent $I$ and $Q$ value is summed with the previous respective sum. As an example, an IPA position calibration using 11-sample integration is shown in Fig.~\ref{fig:f4IPA1CalQPrime}. 

Representative position and angle calibration constants for the three dipole BPMs, calculated using 11-sample integration, are presented in Tab.~\ref{t:AngFit}. It can be seen that IPA and IPB have similar sensitivities, whereas IPC has a lower sensitivity; this is due to a minor fabrication difference.

\setlength{\tabcolsep}{12pt}
\begin{table*}[htbp]
\begin{center}
\begin{tabular}{c c c }\toprule \\[-0.8em]
BPM & Position calibration const. & Angle calibration const.
\\\\[-0.9em]\hline \\[-0.8em]
IPA & $0.184\pm0.002$~${\upmu\mathrm{m}}^{-1}$ & $0.277\pm0.003$~m${\mathrm{rad}}^{-1}$  \\
IPB & $0.168\pm0.002$~${\upmu\mathrm{m}}^{-1}$ & $0.253\pm0.003$~m${\mathrm{rad}}^{-1}$ \\
IPC & $-0.110\pm0.001$~${\upmu\mathrm{m}}^{-1}$ & $-0.157\pm0.002$~m${\mathrm{rad}}^{-1}$\\ \toprule
\end{tabular}
\end{center}
\caption[Position and angular calibration constants]{BPM position and angle calibration constants.}
\label{t:AngFit}
\end{table*}

\section{Cavity BPM system position resolution}
\label{sec:resolution}

The resolution of the BPM system was evaluated using measurements of the bunch trajectory at all three dipole BPMs. Since the bunch follows a straight-line trajectory which can be characterized with measurements from only two BPMs, measurements from the third BPM can be used to estimate the resolution of the system. 

The beam position at BPM $i$, $y_i$, can be represented as a linear combination of the positions of the beam at the other two BPMs, $y_j$ and $y_k$:
\begin{equation}\label{ResPredLinearRep2}
y_{i} = A_{ij} y_{j} + A_{ik} y_{k},
\end{equation}
where $A_{ij}$ and $A_{ik}$ are `geometric' coefficients defined by the relative separations of the three BPMs (Fig.~\ref{fig:Submovers}). The predicted beam position at BPM $i$, $y_{i}^{\mathrm{pred}}$, can therefore be written in terms of the measured positions at BPMs $j$ and $k$:
\begin{equation}\label{ResPredLinearRep}
y_{i}^{\mathrm{pred}} = A_{ij} y_{j}^{\mathrm{meas}} + A_{ik} y_{k}^{\mathrm{meas}}.
\end{equation}
The difference between this and the measured position, $y_{i}^{\mathrm{meas}}$, yields a residual. Under the assumption that all three BPMs have the same resolution, $\sigma_\mathrm{res.}$, the resolution is derived from the standard deviation of the distribution of residuals measured over a batch of sequential beam pulses:
\begin{equation}\label{ResAssumeAllSame}
\sigma_\mathrm{res.} = \textup{std} \bigg\{\frac{(y_i^\mathrm{meas}-y_i^\mathrm{pred})}{\sqrt{1+A_{ij}^2 + A_{ik}^2}}\bigg\}_{ijk}.
\end{equation}

\begin{figure}[htbp]
\centering
\includegraphics[width=1\linewidth]{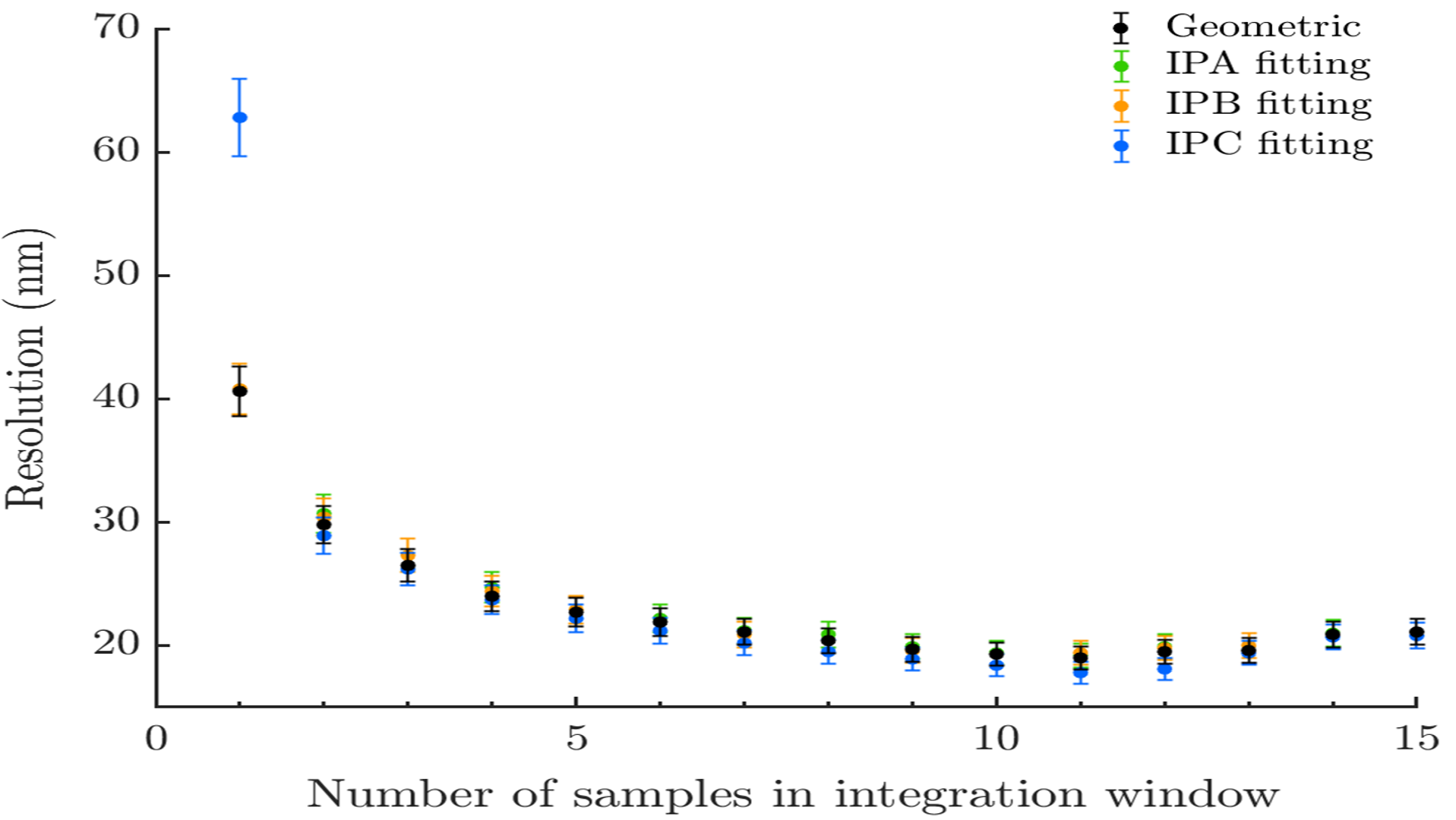}
\caption{Resolution vs. number of samples integrated; the location of each integration window was chosen so as to optimize the geometric resolution. Results for the `geometric' (black) and `fitting' method for each BPM (green, gold, blue) are shown. The error bars represent the statistical uncertainty on the resolution. }
\label{fig:3ResolutionAsFnSampleWindow}
\end{figure}

Detailed studies of the experimental setup to optimize the resolution, including the BPM alignment procedure, are given in~\cite{RamjiawanThesis}. For a data set with bunch charge $0.5\times10^{10}~e$, Fig.~\ref{fig:3ResolutionAsFnSampleWindow} shows the resolution as a function of the number of $I$ and $Q$ samples integrated in real time for the position calculation. It can be seen that the resolution improves from 41~nm (single sample) to an optimal value of 19~nm with 11 samples. No improvement is seen by integrating additional later samples as the BPM waveforms have decayed and the signal levels are low. 

As a cross-check, an alternative, `fitting', method was employed. Here the coefficients $A_{ij}$ and $A_{ik}$ (Eq.~\ref{ResPredLinearRep2}) are fitted to the measured position data set so as to minimise empirically the resolution (Eq.~\ref{ResAssumeAllSame}). 
The fitting method may be applied separately to each of the three BPMs, giving three correlated estimates of the resolution. Were the resolution effectively degraded via the influence of uncontrolled correlated parameters, the empirical fit could yield an improvement over the geometric method~\cite{RamjiawanThesis}. The fitted resolution results for the same data set are also given in Fig.~\ref{fig:3ResolutionAsFnSampleWindow}; the results are in good agreement with the geometric method and confirm that the real-time resolution of 19~nm is the best that could be obtained for these BPMs with the given beam conditions. The resolution results are summarised in Table~\ref{t:IntegratedResolutionImprovement}. 

\begin{table}[htbp]
\newcolumntype{Y}{>{\centering\arraybackslash}X}
\begin{tabularx}{\linewidth}{X Y Y}\toprule
Resolution & Single-sample (nm) & 11-sample (nm)\\\hline
Geometric & $40.6\pm1.0$  & $19.0\pm0.4$\\
IPA fitting & $40.6\pm1.0$ &  $19.2\pm0.6$ \\
IPB fitting & $40.8\pm1.0$  & $19.4\pm0.6$\\
IPC fitting & $62.8\pm1.3$ & $17.6\pm0.4$ \\\toprule
\end{tabularx}
\caption[Single-sample and integrated-sample resolution]{The best single-sample and integrated-sample resolution measurements for the geometric and fitting methods.}
\label{t:IntegratedResolutionImprovement}
\end{table}

\section{ATF2 IP Bunch-by-bunch Feedback System}

\subsection{System design}

\begin{figure}[htbp]
  \centering  
\subfigure[]{\includegraphics[width=0.48\linewidth]{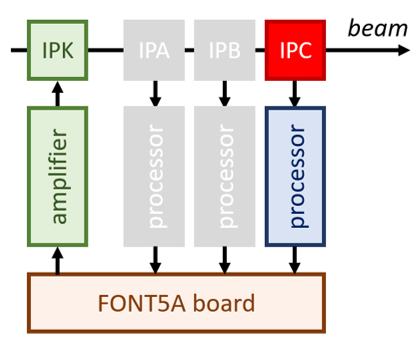}}\hfill
\subfigure[]{\includegraphics[width=0.48\linewidth]{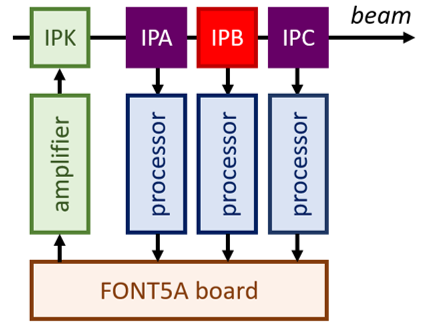}}\hfill
\caption{Diagrams of feedback loops showing dipole cavity BPMs (IPA, IPB and IPC) and stripline kicker (IPK). (a) Single-BPM feedback with beam measurement and stabilization illustrated at IPC (red). (b) Two-BPM feedback, illustrated for position measurements at IPA and IPC (purple) with beam stabilization at IPB (red).}
  \label{fig:FeedbackLoops}
\end{figure}

The high-resolution real-time vertical beam position information from the cavity BPM system was used as input to a closed-loop feedback. For two-bunch trains (see Section~\ref{sec:accel}) the position of the first bunch was measured and used to correct the position of the second bunch. Two feedback operating modes were used, represented functionally in Fig.~\ref{fig:FeedbackLoops}. In single-BPM mode (Fig.~\ref{fig:FeedbackLoops}(a)) the position signal from one BPM was used to derive the correction signal supplied to the kicker IPK, such that the vertical beam position was stabilised at the chosen BPM. For this mode the IP was moved longitudinally from the nominal IP to the center of the chosen BPM so as to directly stabilise the vertical position there. In two-BPM mode (Fig.~\ref{fig:FeedbackLoops}(b)) the IP was placed longitudinally at one BPM; the position signals from the other two BPMs were used to derive a correction signal such that the nano-beam was stabilised vertically at the chosen BPM, which hence served as an independent witness of both the corrected and uncorrected beam positions. Four multiplexers within the firmware allow selection among the three BPMs for their input either individually (mode (a)) or as a pair (mode (b)).

For both modes the correction signal to the kicker is output from the FONT5A board (Fig.~\ref{fig:4FONTBoard}) via a Linear Technology 14-bit LTC2624~\cite{LinearTechnology} digital-to-analogue converter (DAC).

\subsection{Kicker and kicker amplifier}
\label{StriplineKickers}

\begin{figure}[htbp]
\centering
\includegraphics[width=0.8\linewidth]{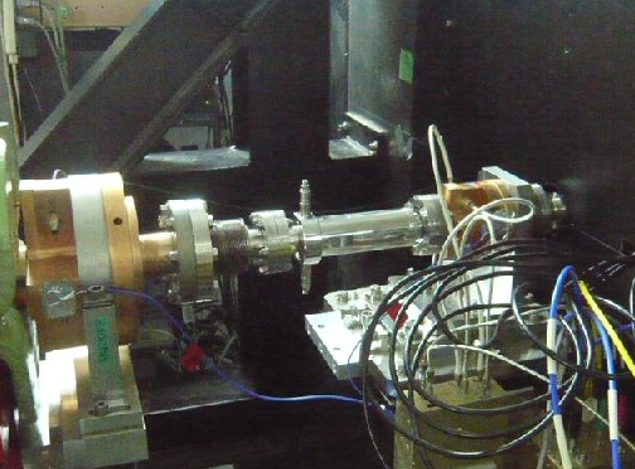}
\caption{Photograph of the IP stripline kicker.}
\label{fig:stripline}
\end{figure}

The correction signal from the FONT5A board requires amplifying before it can be used to drive the kicker (see Fig.~\ref{fig:FeedbackLoops}). The stripline kicker (Fig.~\ref{fig:stripline}) is a modified stripline BPM~\cite{PhysRevSTAB.18.032803} and consists of two conducting strips, $\sim$\SI{12.5}{\cm} in length and separated by \SI{24}{\mm}, at the top and bottom of the inside of the beam-pipe. The custom-made kicker amplifier (see eg.~\cite{ConstanceThesis}) was manufactured by TMD Technologies Ltd~\cite{TMD}. In order to meet the low-latency requirements the amplifier was designed with a fast rise time of 35~ns to reach 90\% of the peak output. The amplifier is capable of providing a drive current of $\pm$30~A.

\subsection{Feedback calculation}
\label{feedback_calculation}

For either feedback mode, the signal sent to the kicker, $V$ (DAC counts), is derived from the measured position offset at the chosen BPM(s) as
\begin{equation}\label{FBAlgorithm}
V=-G\frac{y}{M}+c,
\end{equation}
where $G$ is the feedback gain, $M$ (\SI{}{\um}/DAC counts) is the kicker response calibration constant, and $c$ is an arbitrary offset. If the beam is being stabilized at a location in between BPMs, $y$ refers to the interpolated position. For the case in which the vertical positions of bunches 1 and 2 are 100\% correlated, optimal stabilization of bunch-2 is obtained with $G$ set to unity. For the case of uncorrelated position components $G$ can be adjusted empirically so as to achieve optimal stabilization of bunch-2. The value of $c$ can be controlled via the firmware settings so as to place the stabilised bunch-2 at any desired vertical position within the feedback dynamic range. 

The firmware is designed so that the kicker drive signal is output at the same time relative to the beam arrival regardless of the number of samples integrated in the digital signal processing, up to a maximum of 15 samples. The firmware is also set up to allow a selectable constant kicker drive signal from the DAC with the same timing structure as for a real feedback pulse; this feature is used to evaluate $M$ directly by measuring the response of the beam as a function of the kicker drive signal. 

\subsection{Latency Measurement}

\begin{figure}[hbtp]
\centering
\includegraphics[width=0.8\linewidth]{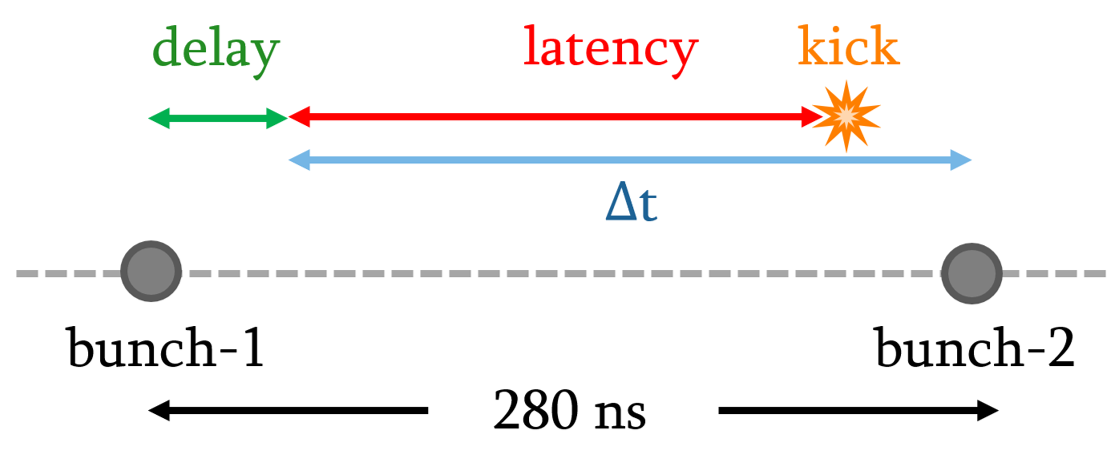}
\caption{Schematic illustrating the principle of the direct latency measurement by adding a controlled delay to the kicker drive output signal.}
\label{fig:4LatencyDesc}
\end{figure}

\begin{figure}[hbtp]
\centering
\includegraphics[width=1\linewidth]{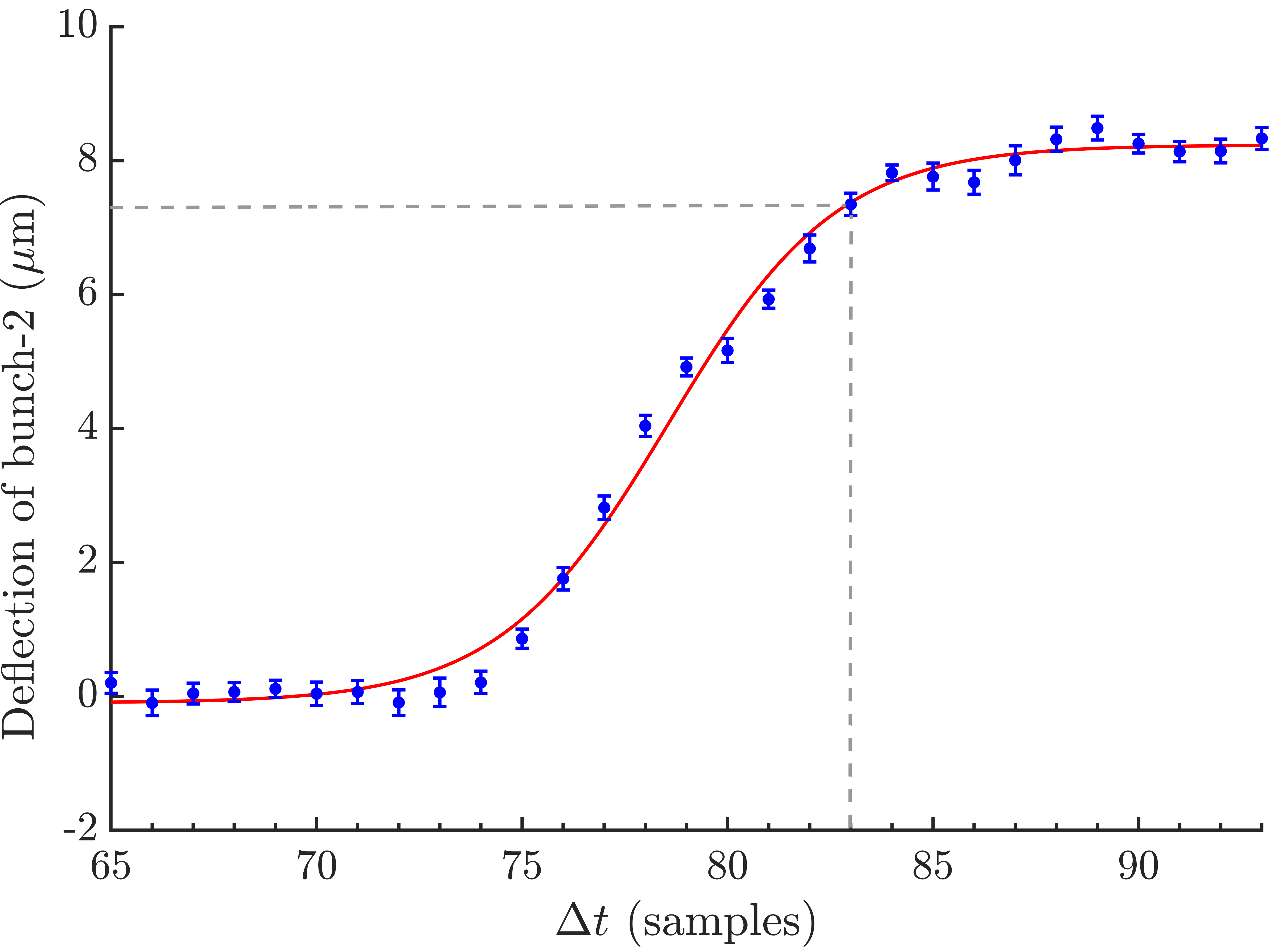}
\caption[Bunch deflection versus feedback signal delay]{The deflection of bunch-2 as a function of $\Delta t$ (\# 2.8 ns samples) defined in Fig.~\ref{fig:4LatencyDesc}. The red line shows a sigmoid fit of the form 
$f(\Delta t)=p_1+\frac{p_2-p_1}{1+10^{p_3-p_4 \Delta t}}$,
where $p_1$, $p_2$, $p_3$ and $p_4$ are fit parameters. The dashed lines show the latency definition at 90\% of maximum deflection.}
\label{fig:4Latency}
\end{figure}

The closed-loop feedback latency is defined as the time interval between bunch-1 passing through the longitudinal center of IPK and the derived kicker correction pulse (for bunch-2) reaching 90\% of its final output value. The latency was measured directly with the beam by adding a controlled delay to a constant kicker drive signal (of 2000 DAC counts) and measuring the resulting position deflection of the second bunch. The principle is illustrated in Fig.~\ref{fig:4LatencyDesc}. For large added delay (small $\Delta t$) the kick arrives too late and bunch-2 is undeflected. For small added delay (large $\Delta t$) the kick arrives in time to fully deflect bunch-2. When the kick arrives in time to kick the bunch by 90\% of the maximum value, then $\Delta t$ is equal to the latency. Sequential triggers were toggled between feedback `off' and `on' to allow running baseline subtraction. 
Fig.~\ref{fig:4Latency} shows the beam deflection as a function of $\Delta t$ from which the latency is measured to be 83 samples, i.e. 232~ns. 

\section{IP Feedback System Performance}

\subsection{Accelerator and feedback setup}
\label{sec:accel}

For the operation of the IP bunch-by-bunch feedback system, the ATF DR was configured to deliver two-bunch trains to ATF2 with a bunch separation of \SI{280}{ns}. The train repetition rate was 1.56~Hz. This setup provides a high degree of correlation between the vertical positions of the two bunches in each train~\cite{NevenThesis}, which yields the conditions for optimal feedback performance in stabilizing the second bunch. 

The limited dynamic range of the dipole BPMs for optimal resolution necessitates both their good transverse centering w.r.t.\ the beam trajectory and, ideally, small beam jitter at each BPM. Each final-focus-system quadrupole is mounted on transverse movers, which allows for adjustments to both the incoming beam position and angle. In particular, moving QD0FF (see Figure~\ref{fig:ATFDiagram}) vertically adjusts the vertical IP position, while moving QF1FF (or the upstream QF7FF) adjusts the vertical beam incoming angle~\cite{BeamWaist2, BeamWaist1}. The beam trajectory is first globally aligned with the electrical centers of the IP BPMs, and fine adjustments are then made to center each BPM w.r.t.\ the beam by using the BPM movers (Fig.~\ref{fig:Submovers}).

For single-BPM feedback (Fig.~\ref{fig:FeedbackLoops}(a)), small beam jitter is achieved by setting the IP at the longitudinal center of the feedback BPM~\cite{Transients}. For two-BPM feedback (Fig.~\ref{fig:FeedbackLoops}(b)) the situation is more difficult as the extreme IP angular divergence produces increasingly large beam jitter as longitudinal distance from the IP increases. 
With the nominal optics configuration the jitter at the feedback BPMs can exceed the dynamic range for best resolution. Therefore, for two-BPM feedback operation an optics configuration with a reduced angular divergence at the IP was used. This yields a reduced beam jitter at the feedback BPMs, although at the expense of increasing the IP beam jitter. These optics are designed such that the ATF2 beamline has the same magnet strength as for the nominal optics except within the matching section. With these optics, the vertical $\beta$-function at the IP is \SI{12}{\cm}.

The BPMs were set up for optimal performance, and calibrated, as described in Section~\ref{sec:position}. In order to make a direct comparison between the data with feedback `on' and `off' within a given dataset, the feedback was toggled between on and off on alternate bunch trains. 

\subsection{Single-BPM IP Feedback Results}

\begin{figure}[!bhtp]
\centering
\includegraphics[width=1\linewidth]{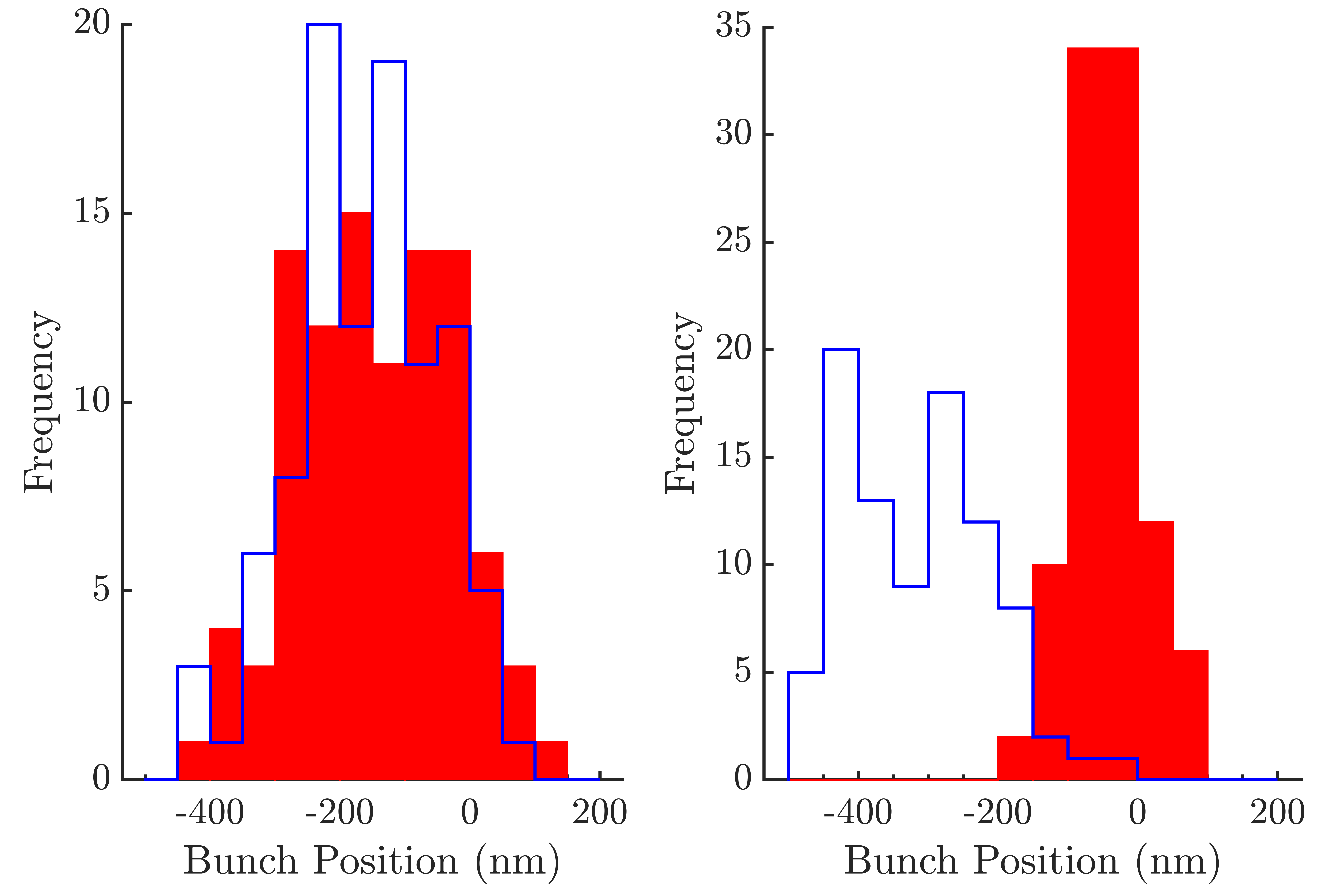}
\caption{Distributions of bunch positions measured at IPC, for bunch-1 (left) and bunch-2 (right) with feedback off (blue) and feedback on (red).}
\label{fig:50nmHist}
\end{figure}

\begin{figure}[htbp]
  \centering
  \subfigure[]{\includegraphics[width=0.4\textwidth]{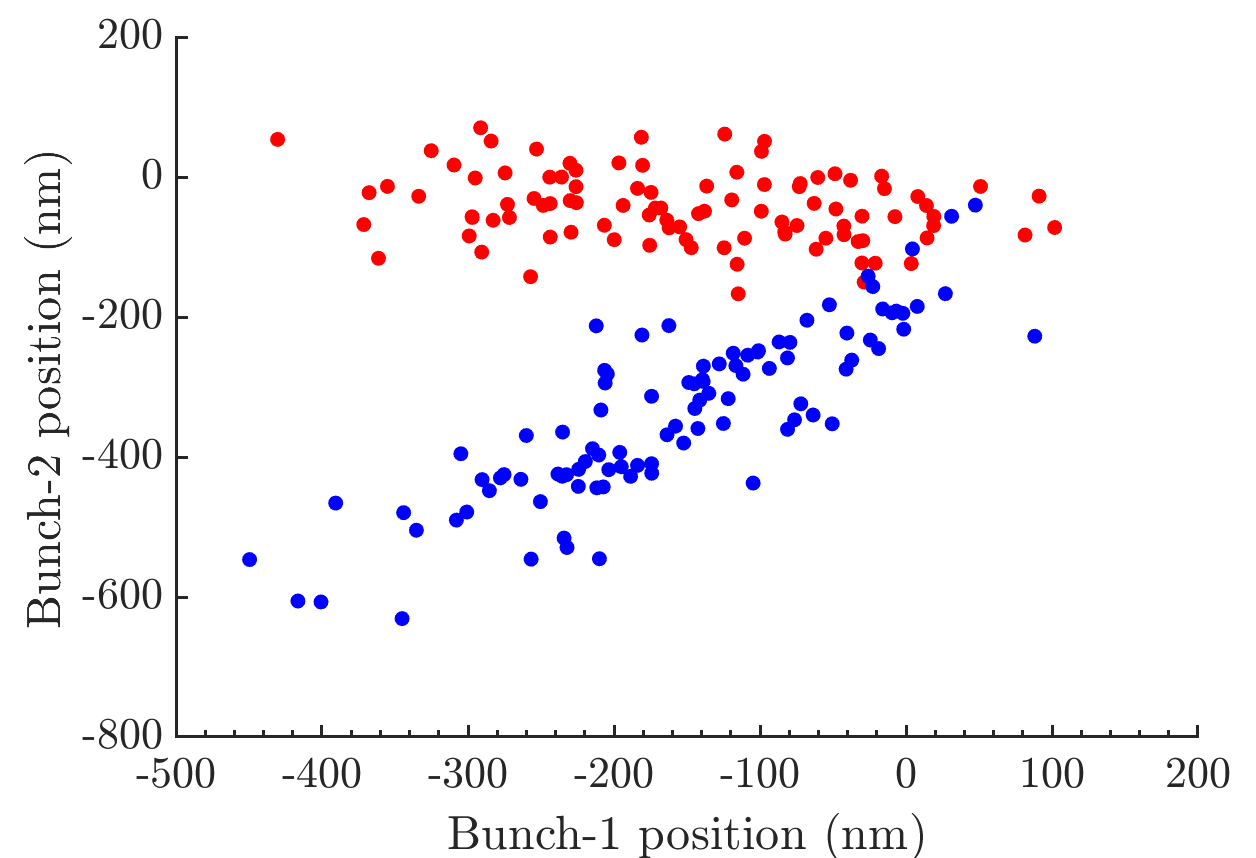}}
\hfill
  \subfigure[]{\includegraphics[width=0.4\textwidth]{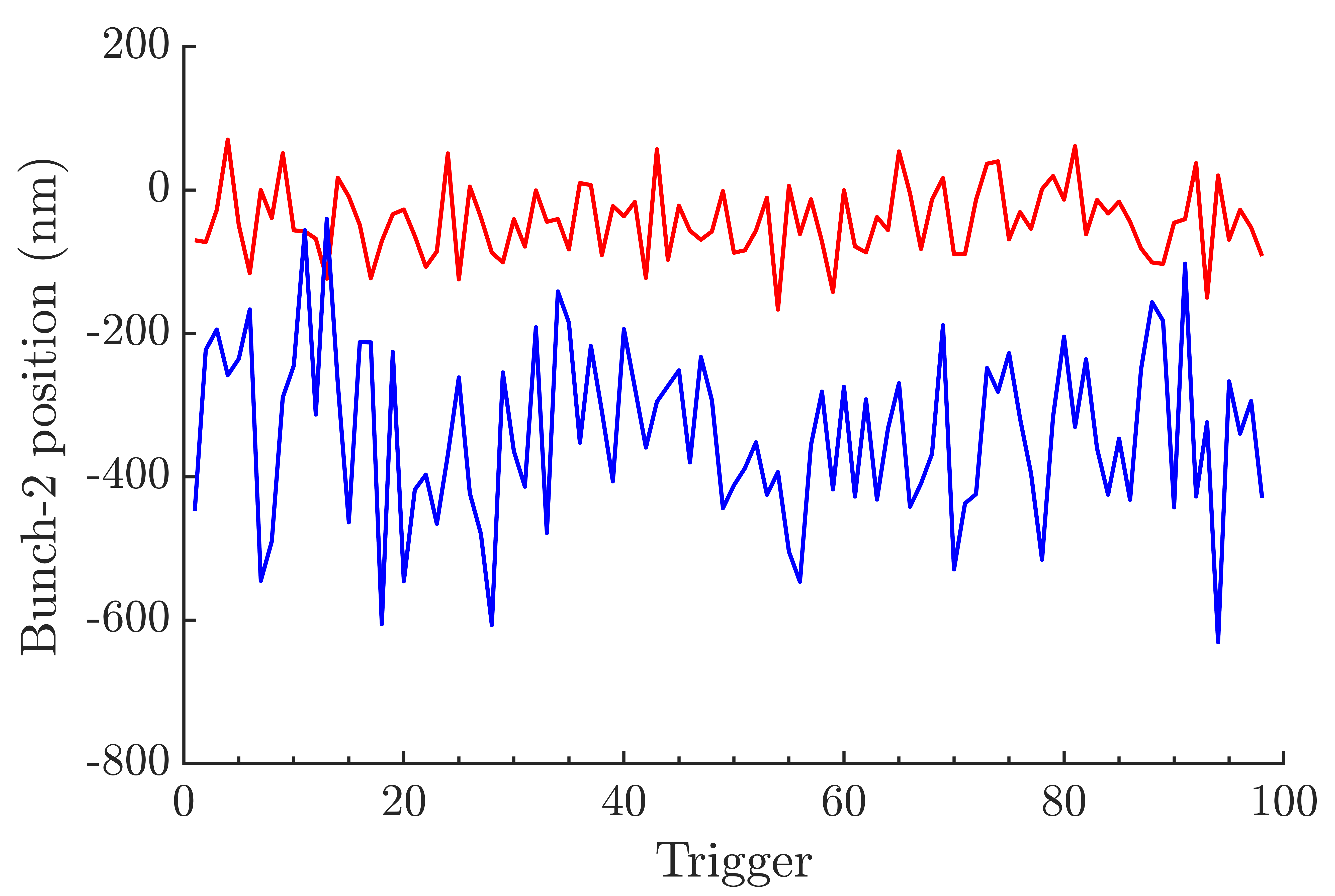}}
  \caption{(a) IPC bunch-2 position versus bunch-1 position and (b) bunch-2 position versus trigger number; for feedback off (blue) and feedback on (red).}
\label{fig:51Correl}
\end{figure}

Single-BPM feedback was operated with a bunch charge of $0.8\times10^{10}e^-$, with the IP set at IPC. 
The feedback gain was set to 0.8 to account for the imperfect bunch-to-bunch position correlation, as determined from correlation measurements taken at the start of the shift (Table~\ref{tab:50nmHist}). Further analysis has suggested, however, that the correlation decreased during the shift. A 10-sample integration window was found empirically to optimize the resolution.
The feedback performance is illustrated in Fig.~\ref{fig:50nmHist} and summarised in Table~\ref{tab:50nmHist}, where we compare feedback-on and feedback-off results. Since bunch-1 provides the input to the feedback its position is unaffected by the correction. By contrast, the bunch-2 mean position is zeroed by the feedback and its jitter is substantially reduced, from \SI{119}{\nm} to \SI{50}{\nm}. 
The same dataset is used in Fig.~\ref{fig:51Correl}, which shows the effect of the feedback on the bunch-to-bunch correlation as well as on the time-sequence of the bunch-2 position.

\begin{table}[htbp]
 \caption{Position jitters and bunch-to-bunch position correlation with feedback off and on, for single-BPM feedback.}
\newcolumntype{Y}{>{\centering\arraybackslash}X}

\newcolumntype{M}{>{\centering\arraybackslash}m{0.8cm}}
\begin{tabularx}{1\linewidth}{M Y Y Y}
\toprule
\textbf{FB} &     \multicolumn{2}{c}{\textbf{Position jitter}} &  \textbf{Correlation}\\
        &     \multicolumn{2}{c}{{(nm)}} &  {(\%)}
      \\ \hline
& \textbf{Bunch-1} & \textbf{Bunch-2} & 
 \\\hline
 Off & 109 $\pm$ 11 & 119 $\pm$ 12  & $85.1^{+2.5}_{-3.5}$\\  
 
 On & 118 $\pm$ 12  &  50 $\pm$ 5 & $-26.0^{+9.8}_{-8.8}$\\
       \toprule
   \end{tabularx}
   
   \label{tab:50nmHist}
\end{table}

The expected level of beam stabilization can be computed from the bunch jitter and the incoming bunch-to-bunch correlation. The corrected bunch-2 position, $Y_2$, in terms of the uncorrected bunch-1 and bunch-2 positions, $y_1$ and $y_2$, respectively, is
\begin{equation}
Y_2=y_2- G y_1+c.
\label{4CorrectedPos}
\end{equation}
Taking the variance of Eq.~\ref{4CorrectedPos} gives
\begin{equation}
\label{4FBPred}
\sigma_{Y_{\mathrm{2}}}^2 = \sigma_{y_{\mathrm{1}}}^2 + G\sigma_{y_{\mathrm{2}}}^2 - 2G\sigma_{y_{\mathrm{1}}}\sigma_{y_{\mathrm{2}}}\rho_{12},
\end{equation}
where $\rho_{12}$ is the bunch-to-bunch correlation and $\sigma_{Y_\mathrm{2}}$, $\sigma_{y_\mathrm{1}}$ and $\sigma_{y_\mathrm{2}}$ represent the jitters on positions $Y_\mathrm{2}$, $y_\mathrm{1}$ and $y_\mathrm{2}$, respectively. 

The measured incoming position correlation between bunches 1 and 2 (feedback off) is about 85\% (Table~\ref{tab:50nmHist}); hence, from Eq.~\ref{4FBPred}, the expected feedback-corrected jitter for bunch-2 is $65\pm11$ nm, which is in reasonable agreement with the measured performance. With feedback on, the measured correlation between bunches 1 and 2 is $-26.0^{+9.8}_{-8.8}$ (Table~\ref{tab:50nmHist}), which implies a slight over-correction. This naively suggests that an improved feedback-corrected jitter would have been possible with a slightly lower gain. Limited beam operation availability at the facility did not allow this to be verified at the time, but it could be investigated in future beam studies.  

The theoretically optimum performance is obtained for 100\% correlation between bunches 1 and 2, i.e. $\rho_{12}=1$, comparable bunch jitters, $\sigma_{y_{\mathrm{1}}}=\sigma_{y_{\mathrm{2}}}$, and feedback gain $G=1$. With these conditions fulfilled, the ultimate limit to stabilization is determined by the BPM resolution, $\sigma_{\mathrm{res.}}$:
\begin{equation}
\label{eq:resolution}
\sigma_{Y_{\mathrm{2}}}=\sqrt{2}\sigma_{\mathrm{res.}}.
\end{equation}
For a real-time BPM resolution of c. \SI{19}{\nm} (Section~\ref{sec:resolution}) the ultimate feedback performance in single-BPM mode (Eq.~\ref{eq:resolution}) would hence be stabilization of bunch-2 to c. 27~nm, so there is in principle still a margin for improvement of the feedback performance reported here, subject to improved beam conditions.   

\subsection{Two-BPM IP Feedback Results}

Two-BPM feedback was operated with a bunch charge of $0.5\times10^{10}e^-$, with the IP set at IPB and with IPA and IPC used as inputs to the feedback; hence IPB was used as an independent witness of the feedback performance. The longitudinal separations of IPA and IPC from the beam waist yield much larger position signal levels and higher signal-to-noise ratios. The sample window was chosen empirically to optimize the resolution, here with a measured resolution of $\sim$\SI{31.2}{\nm}, for a five-sample window. Feedback was operated with a gain of 0.8 to account for the differences in the position jitter between the two bunches and the imperfect bunch-to-bunch position correlation (Table~\ref{2BPMFBPerformance}).

\begin{figure}[htbp]
\centering
\includegraphics[width=1\linewidth]{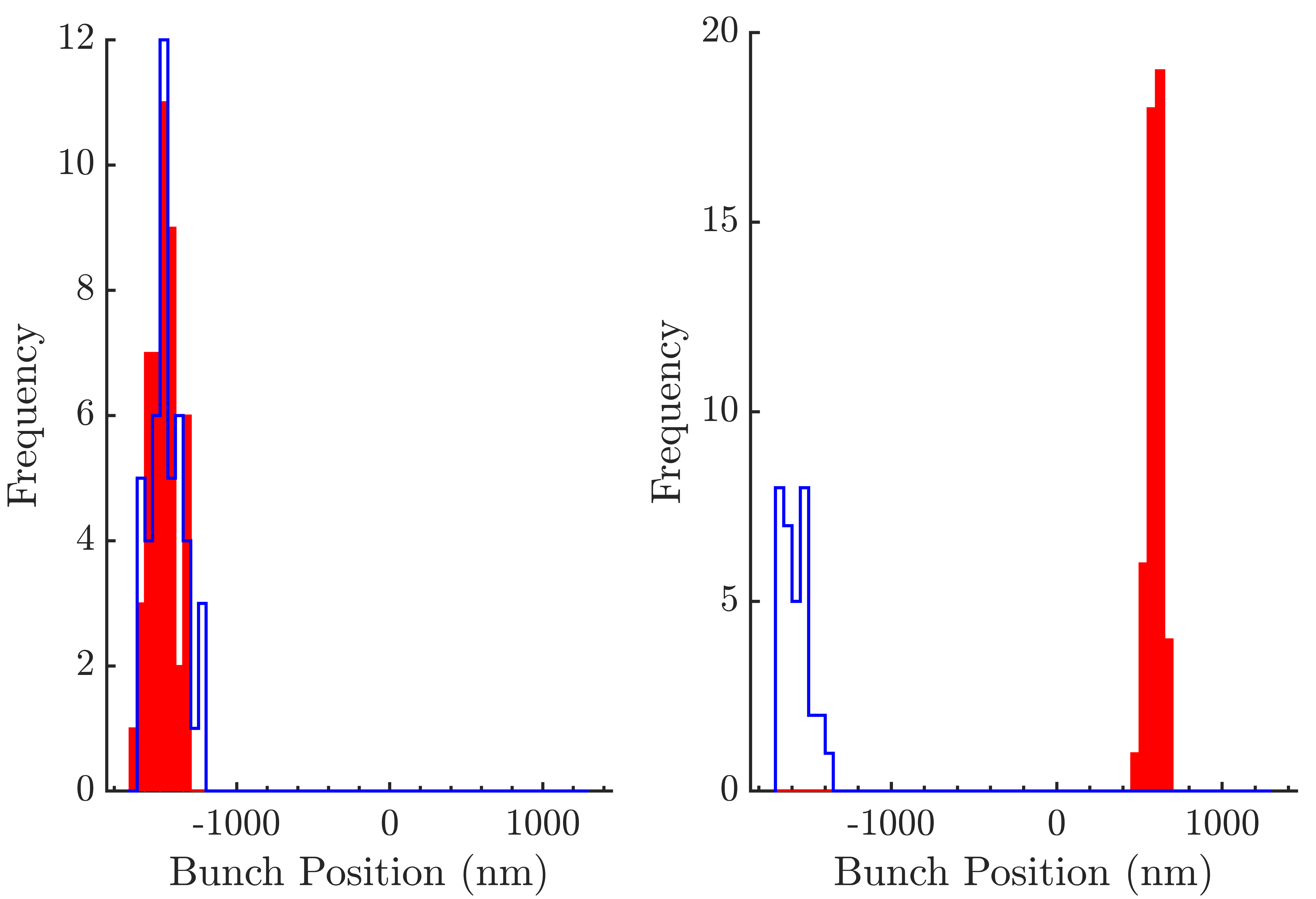}
\caption{Distributions of bunch-1 (left) and bunch-2 (right) positions measured at IPB, with feedback off (blue) and feedback on (red). Feedback was performed in 2-BPM mode, stabilizing at IPB using beam position measurements from IPA and IPC.}
\label{fig:40nmHist2}
\end{figure}

The feedback performance is illustrated in Fig.~\ref{fig:40nmHist2} and summarised in Table~\ref{2BPMFBPerformance}, where we compare feedback-on and feedback-off results. Since bunch-1 provides the input to the feedback its position is unaffected by the correction. By contrast, the bunch-2 jitter is substantially reduced, from 96~nm to 41~nm. The measured incoming position correlation between bunches 1 and 2 (feedback off) is about 92\% (Table~\ref{2BPMFBPerformance}); hence, from Eq.~\ref{4FBPred}, the expected feedback-corrected jitter for bunch-2 is $40\pm11$~nm, which is in excellent agreement with the measured value. 

\begin{table}[htbp]
\newcolumntype{Y}{>{\centering\arraybackslash}X}
\newcolumntype{M}{>{\centering\arraybackslash}m{0.8cm}}
  \centering
   \caption{Position jitters and bunch-to-bunch position correlation with feedback (FB) off and on, for 2-BPM feedback.}
   \begin{tabularx}{1\linewidth}{M Y Y Y}
       \toprule\\[-0.9em]
    \textbf{FB} &     \multicolumn{2}{c}{\textbf{Position jitter}} &  \textbf{Correlation}\\
    & \multicolumn{2}{c}{(nm)} & (\%)
      \\\hline \\[-0.9em]
& \textbf{Bunch-1} & \textbf{Bunch-2} & \\\\[-0.9em] 
 \hline \\[-0.9em]
 Off & 106 $\pm$ 11 & 96 $\pm$ 10  & $91.6^{+1.8}_{-3.1}$ \\  
 
 On & 100 $\pm$ 10  &  41 $\pm$ 4 & $41.3^{+9.1}_{-12.3}$\\\\[-0.9em]
       \toprule
   \end{tabularx}
   \label{2BPMFBPerformance}
\end{table} 
 
In Fig.~~\ref{fig:40nmHist2} it can be seen that the mean corrected bunch-2 position was at $\sim$\SI{0.5}{\um}, which simply arises from the residual relative transverse offsets between IPB and IPA/IPC; if desired this offset can trivially be removed with a compensating constant offset term, shown as $c$ in the feedback algorithm (Eq.~\ref{FBAlgorithm}). 

With feedback on the measured correlation between bunches 1 and 2 is about 41\% (Table~\ref{2BPMFBPerformance}), which implies an under-correction. This suggests that an improved feedback performance would have been possible with a higher gain. As previously noted, at the time, beam operation availability was limited and this could not be verified but it could be confirmed with further beam studies.   

In this feedback mode, with stabilization at IPB, the feedback BPMs, IPA and IPC, contribute position information in the ratio 32:68, determined by their relative distances from IPB (Fig.~\ref{fig:Submovers}). Hence, the theoretically best-possible resolution on the corrected beam position at IPB is given by:
\begin{equation}
\sigma_{\mathrm{IPB}} = \sqrt{0.32^2\sigma_{\mathrm{res.}}^2+0.68^2\sigma_{\mathrm{res.}}^2} = 0.75\sigma_{\mathrm{res.}};
\end{equation}  
i.e. beam stabilization at IPB as low as c. 23~nm would have been achievable in principle given the measured resolution of 31~nm. Correspondingly, with the best achieved resolution of 19~nm, and a perfect feedback correction, stabilization to 15~nm would be theoretically possible. Hence there is still a margin for improvement of the feedback performance reported here, subject to improved beam conditions.   

\section{SUMMARY AND CONCLUSIONS}

We have reported the design, operation and performance of a high-resolution, low-latency, bunch-by-bunch feedback system for beam stabilization. The system includes high-resolution cavity BPMs, two stages of analogue signal down-mixing system, and a digital board incorporating an FPGA. The FPGA firmware allows for the real-time integration of up to fifteen samples of the BPM waveforms so that feedback can be performed within a latency of \SI{232}{\ns}. We have shown that this real-time sample integration improves the beam position resolution, with measured resolutions as good as \SI{19}{\nm}, which consequently improves the feedback performance. 

In~\cite{Inoue} results were reported using similar cavity BPMs, but with a higher design quality factor: data were recorded and the resolution was determined in a subsequent offline analysis using a function that included 10 free parameters to account for uncontrolled effects; a resolution of $\sim$9~nm was thereby obtained. In addition, the position-calibration constant was not measured directly at the most sensitive resolution setting, but was interpolated from measurements made with added signal attenuation, at lower position sensitivity. Furthermore, no attention was paid to signal processing latency as the BPMs were not used for bunch-by-bunch feedback. 

We have made several significant advances since this earlier study: 1) the BPMs were calibrated w.r.t.\ position directly at the most sensitive resolution setting and the respective calibration factors were applied in the subsequent BPM operations; 2) the signal processing was done in real-time and with low latency, so as to permit the BPMs to be used for bunch-by-bunch feedback; 3) the resolution was measured directly, in real-time, without fitting any extra parameters. The high BPM resolution was hence utilised directly for stabilization of the beam, and is not merely an impressive offline performance figure of merit.

The feedback was operated in two complementary modes to stabilise the vertical position of the ultra-small beam produced at the focal point of the ATF2 beamline at KEK. In single-BPM feedback mode, beam stabilization to $50\pm5$~nm was demonstrated. In two-BPM feedback mode, beam stabilization to $41\pm4$~nm was achieved, in good agreement with the predicted value, given the incoming beam conditions, of \SI{40}{\nm}. 

Some margin remains to improve the feedback performance by increasing the degree of bunch-to-bunch position correlation in the incoming beam, and suitably optimising the gain. For the best achieved position resolution to date, and for 100\% bunch-to-bunch correlation, an ultimate beam stabilization to about 15~nm is in principle achievable with the current hardware. Should ATF/ATF2 beam operations resume, this will be the subject of future feedback studies.

\section{ACKNOWLEDGEMENTS}
We thank the KEK ATF staff for their outstanding logistical support and for providing the beam time and the necessary stable operating conditions for this research. In addition, we thank our colleagues from the ATF2 collaboration for their help and support. In particular, we thank the KNU group for fabricating the low-quality-factor BPMs, the LAL group from the Paris-Saclay University for providing the BPM mover system, and the KEK group for making available the analogue signal-processing down-mixing electronics. 

We acknowledge financial support for this research from the United Kingdom Science and Technology Facilities Council via the John Adams Institute, University of Oxford, and CERN, CLIC-UK Collaboration, Contract No. KE1869/DG/CLIC. The research leading to these results has received funding from the European Commission under the Horizon 2020/Marie Sklodowska-Curie Research and Innovation Staff Exchange (RISE) project E-JADE, Grant Agreement No. 645479.


\providecommand {\noopsort}[1]{}\providecommand {\singleletter}[1]{#1}%

\end{document}